\documentclass[aps,twocolumn,nofootinbib,longbibliography,notitlepage]{revtex4-1}
\usepackage{amsmath}
\usepackage{amstext}
\usepackage{amssymb}
\usepackage{xfrac}
\usepackage[colorlinks,citecolor=blue]{hyperref}
\usepackage{graphicx}
\usepackage{amsmath}
\usepackage{amstext}
\usepackage{amssymb}
\usepackage{amsfonts}
\usepackage{longtable,booktabs}
\usepackage{hyperref}
\usepackage{url}
\usepackage{subfigure}
\usepackage{dsfont}
\usepackage{booktabs}
\usepackage{amsbsy}
\usepackage{dcolumn}
\usepackage{amsthm}
\usepackage{bm}
\usepackage{esint}
\usepackage{multirow}
\usepackage{hyperref}
\usepackage{cleveref}\usepackage{amsmath}
\usepackage{amsfonts}
\usepackage{amssymb}
\usepackage{mathrsfs}
\usepackage{amsfonts}
\usepackage{amsbsy}
\usepackage{dcolumn}
\usepackage{bm}
\usepackage{multirow}
\usepackage{color}
\usepackage{extarrows}
\usepackage{datetime}
\usepackage{comment}
\usepackage[super]{nth}
\usepackage{tikz-cd}

\hypersetup{
    colorlinks=magenta,
    linkcolor=blue,
    filecolor=magenta,
        urlcolor=magenta,
}
\def\Z{\mathbb{Z}}

\newcommand{\ket}{\rangle}

\def\Z{\mathbb{Z}}

\usepackage{extarrows}

\usepackage{datetime}

\usepackage{comment}

\usepackage{lineno}

\begin{document}
	
	
	\title{Fracton physics of spatially extended excitations. II.  Polynomial ground state degeneracy of exactly solvable models}
	
	
	
	\author{Meng-Yuan Li}
	\affiliation{School of Physics and State Key Laboratory of Optoelectronic Materials and Technologies, Sun Yat-sen University, Guangzhou, 510275, China}
	
	\author{Peng Ye}\email{yepeng5@mail.sysu.edu.cn}
	\affiliation{School of Physics and State Key Laboratory of Optoelectronic Materials and Technologies, Sun Yat-sen University, Guangzhou, 510275, China}

 	\date{\today}

\begin{abstract}
Generally, ``fracton''  topological orders are referred to as gapped   phases that support \textit{point-like topological excitations} whose mobility is, to some extent, restricted. In our previous work [\href{https://doi.org/10.1103/PhysRevB.101.245134}{Phys. Rev. B 101, 245134 (2020)}], a large class of   exactly solvable models on hypercubic lattices are constructed. In these models,  \textit{spatially extended excitations} possess generalized fracton-like properties:  not only mobility but also deformability is restricted.  As a series  work, in this paper, we proceed further to   compute ground state degeneracy (GSD)    in both isotropic and anisotropic lattices. We decompose and reconstruct  ground states  through a consistent collection of subsystem ground state sectors, in which mathematical game ``coloring method'' is applied. Finally, we are able to systematically obtain GSD formulas (expressed as $\log_2 GSD$) which exhibit diverse kinds of polynomial dependence on   system sizes.    For example, the GSD of the model labeled as $[0,1,2,4]$ in four dimensional isotropic hypercubic lattice  shows   $ 12L^2-12L+4$ dependence on the linear size $L$ of the lattice.  Inspired by existing results [\href{https://doi.org/10.1103/PhysRevX.8.031051}{Phys. Rev. X 8, 031051 (2018)}], we expect that the   polynomial formulas   encode      geometrical and topological fingerprints of higher-dimensional manifolds beyond toric manifolds used in this work. This is left to future investigation. 

 \end{abstract}

 \maketitle

\tableofcontents


\section{Introduction}
\label{sec:intro}

 Topological orders are gapped phases of matter that cannot be characterized by symmetry-breaking order parameters and are robust without the need of symmetry protection   \cite{Wen2016RMPzoo}. To characterize topological orders, topological order  parameters, such as braiding statistics and ground state degeneracy (GSD)   are applied to characterize topological orders, as long as the bulk gap is not closed   \cite{PhysRevB.40.7387,wen_stacking}.
Recently, a new kind of orders, dubbed \textit{fracton (topological) order}, has been drawing much attention   \cite{Nandkishore2019,2020Fracton,Nandkishore2019,Vijay2015,Vijay2016,Prem2017,Chamon05,Vijay2015,Shirley2019,Ma2017,Haah2011,Bulmash2019,Prem2019,Slagle2017,Shirley2018a,Slagle2019a,Shirley2018,2018PhRvB..97h5116P,Pai2019a,Sala2019,Pretko2018,Pretko2017,Ma2018,Pretko2017a,Radzihovsky2019,Dua2019,PhysRevLett.122.076403,PhysRevX.9.031035,2019arXiv190404815K,Pretko18localization,PhysRevB.100.125150,PhysRevB.99.245135,PhysRevB.97.144106,PhysRevB.99.155118,MaHigherRankDQC,PhysRevLett.126.101603, 2021PhRvB.103o5140W, 2020ScPP....9...73G, 2020ScPP....9...76N, 2020PhRvB.102t5106P, 2020arXiv200407251W, 2020PhRvR...2c3331S,2020arXiv200400015S, 2020PhRvR2c3124G, 2020arXiv200212026S, 2020PhRvR...2d3165A, 2020PhRvR...2c3300W, 2020arXiv200105937P,ye19a,FS1,PhysRevResearch.3.013226,ye21RG,2020PhRvR...2d3219W}. Fracton orders exhibit an interesting interplay of topology and geometry in quantum many-body physics. One of remarkable topological properties is the restricted mobility of topological excitations. More precisely,   topological excitations in disorder-free fracton order systems cannot be moved away from their initial locations by local operators or more general local noise. In the extreme case where mobility towards all spatial directions is  entirely   frozen, such a kind of topological excitations is dubbed ``fracton'' following   nomenclature widely used in the literature. The remaining excitations are ``subdimensional particles'' that have partial mobility and are thus movable within certain subspaces. 
Such a restriction on mobility of excitations is essentially rooted in topological reasons, which becomes crystal clear in exactly solvable model construction in terms of stabilizer codes  \cite{Haah2011}.  Recently,  there have been multidisciplinary research activities in quantum information, condensed matter physics, and high energy physics, e.g., fragmentation of Hilbert space, robust storage of quantum memory, new duality, gravity, and higher rank gauge theory. 
\begin{table*}[t]
	\caption{ \textbf{Ground state degeneracy (GSD) results  of   $[d_n,d_s,d_l,D]$ models on both isotropic and anisotropic lattices with periodic boundary conditions (PBC).} GSD as a function of $L$ is computed in isotropic lattices with linear size $L$.  Otherwise, GSD is computed in anisotropic lattices, where $L_n$ is the linear size along the $\hat{x}_n$ direction of   lattices. It is known that in X-cube model, the coefficients of the linear terms are the first Betti numbers of leaves. Therefore, these coefficients reflect the topological properties of leaves. While as we can see, in $[0,1,2,D]$ models with $D\geq 4$, except for the linear term, we also obtain terms of higher degrees. Moreover, in anisotropic models, there are also crossing terms between sizes along different directions. Such terms of higher degrees call for further investigation on their relations to  topological and geometric properties of the models.}
	\label{table_result}
	\begin{ruledtabular}
		\begin{tabular}{p{90pt} p{330pt}}
			Model& $\log_2 GSD$\\
			\colrule
			X-cube & $6L-3$ \\
			$[0,1,2,4]$ & $12L^2-12L+4$ \\
			$[1,2,3,4]$ & $12L-6$ \\
			$[0,1,2,D]$ & $\sum_{n=0}^{D-2} D \times   \binom{D-1}{n} (-1)^{D+n} L^n$ \\
			$[D-3,D-2,D-1,D]$ & $\binom{D}{D-2} \times (2L-1)$ \\
			\hline
			
			 \hline
			X-cube & $2L_1 + 2L_2 + 2L_3 -3$ \\
			$[0,1,2,4]$ & $2\sum_{i<j} L_i L_j  - 3\sum_i L_i + 4$ \\
			$[1,2,3,4]$ & $3 L_1+3L_2 +3L_3 +3L_4-6$ \\
			$[0,1,2,5]$ & $2 \sum_{i<j<k} L_i L_j L_k - 3\sum_{i<j} L_i L_j + 4\sum_i L_i -5$ \\
			$[2,3,4,5]$ & $4L_1 + 4L_2 + 4L_3+ 4L_4 + 4L_5 -10$\\
		\end{tabular}
	\end{ruledtabular}
\end{table*}
 
In the literature, most of exactly solvable models of fracton orders only contain   \textit{point-like} topological excitations with restricted mobility. Nevertheless, in 3D and higher-dimensional pure topological orders,   we have been witnesses to  ongoing research progress  on fruitful physics of spatially extended topological excitations, such as strings and membranes   \cite{lantian3dto1,lantian3dto2,yp18prl,ypdw,wang_levin1,jian_qi_14,string5,PhysRevX.6.021015,string6,ye16a,YeGu2015,corbodism3,YW13a,ye16_set,2018arXiv180101638N,2016arXiv161008645Y,string4,PhysRevLett.114.031601,3loop_ryu,WANG2020135516,2016arXiv161209298P,Ye:2017aa,Tiwari:2016aa,2012FrPhy...7..150W,PhysRevResearch.3.023132,Zhang:2021ycl}. Therefore, it is fundamentally  important to explore fracton-type physics of spatially extended excitations.   For this goal, in the  work  \cite{ye19a}, we proposed a series of exactly solvable lattice models that are uniquely labeled by a group of four integers: $[d_n,d_s,d_l,D]$. Here, $D$ means $D$-dimensional cubic lattice.    To our surprise, these models contain not only point-like subdimensional excitations, but also spatially extended excitations   whose  mobility and deformability are restricted to some extent. Besides,   we   construct spatially extended excitations with stable non-manifold-like shapes, which are termed ``complex excitations''. Excitations with stable disconnected shapes are also discussed.   In conclusion, we classify   topological excitations   into $4$ sectors, which are respectively trivial, simple, disconnected and complex excitations.

 In conventional topological orders,  GSD depends on topology of base manifolds where   many-body systems are spatially defined, and cannot be lifted by local operators in the thermodynamical limit.  In fracton orders,  GSD is also robust against local operators and ground states are indistinguishable from each other via local measurements. But GSD of fracton orders is no longer uniquely determined by topology of base manifolds. Rather, GSD quantitatively depends on various topological and geometric properties, such as foliation and boundary conditions  \cite{Vijay2015,2018PhRvB..97h5116P,Shirley2019,Shirley2018,Vijay2016,Ma2017,2020AnPhy.42268318Q,2017arXiv170100762V,Haah2011,PhysRevB.97.125101}. For example, GSD of fracton orders in the X-cube model exponentially explodes with respect to   the linear size $L$ of the system, i.e., $GSD\sim 2^{6L-3}$. In the literature, the coefficients of linear terms of $\log_2 GSD$ have been identified as the first Betti numbers with $\Z_2$ coefficients of leaves  \cite{Shirley2018}, which opens a new condensed-matter window into   mathematics. In short, GSD is a very important topological order parameter, and also a significant character that explicitly distinguishes fracton orders from conventional topological orders.

As a series work, in this paper,  we move forward and    study GSD  of $[d_n,d_s,d_l,D]$ models  constructed in Ref.~\cite{ye19a}.  As an initial attempt, we present a combinatorial method to rigorously derive GSD formulas of a subset of $[d_n,d_s,d_l,D]$ models, which are  summarized in Table.~\ref{table_result}. Our results show that   GSD  values are expressed as diverse  kinds of polynomial dependence on the linear sizes of both isotropic and anisotropic hypercubic lattices. As we shall discuss in the main text, the polynomial expressions of $\log_2 GSD$ of $[0,1,2,D]$ models are fundamentally rooted in exotic multi-level foliation structures of the models ($D\geq 4$). \textit{Here, multi-level foliation means that a ground state  of a $[0,1,2,D]$ model restricted in a $(D-1)$-dimensional subspace is a $[0,1,2,D-1]$ ground state which again exhibits a foliation structure.}  A comprehensive discussion on  restriction of ground states will be given in Sec.~\ref{subsec_res_of_gs}. Inspired by elegant results in Ref.~\cite{Shirley2018},  we expect   these polynomials    may potentially encode rich mathematics of topology and geometry, which will be one of   appealing future directions.

 To obtain GSD formulas, we technically perform an exotic   decomposition of base manifolds for $[d_n,d_s,d_l,D]$ models, which can be intuitively recognized as a class of foliation structures of different dimensions. Accordingly, we  also need to decompose Ising configurations (in spin-$z$ basis), stabilizers and ground states of $[d_n,d_s,d_l,D]$ models. By proving that there is a one-to-one correspondence between a ground state sector in the fracton ordered model and a consistent collection of subsystem ground state sectors, we finally obtain GSD with a combinatorial algorithm. The latter is an interesting math game:  {coloring method}.

  This paper is organized as follows. In Sec.~\ref{sec_pre}, we introduce some useful  preliminary  materials that are critical to our proof and computation.   In Sec.~\ref{subsec_res_of_gs}, we present the details of representing ground states with lower-dimensional data. With the preparations in the former two sections,  Sec.~\ref{subsec:overview} is devoted to the general steps toward the ground state degeneracy. In Sec.~\ref{sec_examples}, we demonstrate the calculation of GSD of some typical $[d_n,d_s,d_l,D]$ models following the general steps. In Sec.~\ref{sec_conc}, we give a brief summary of our computation of GSD and  potential implications, and tentatively discuss about some relevant questions yet to be solved. There are also two useful appendix at the end of the paper.

\section{Preliminaries}
\label{sec_pre}

\subsection{Geometric notations}
\label{subsec_geo_not}

To refer to objects of all dimensions,    it is useful to introduce a series of geometric notations.  We begin with an introduction of a coordinate system. In this paper, as we mainly focus on cubic lattice, we can refer to a \textit{$d$-cube} denoted as $\gamma_d$ via the coordinate of its geometric center. Here a $d$-cube is a $d$-dimensional analog of a cube\footnote{For example, $0$-cubes are vertices, $1$-cubes are links, and $2$-cubes are plaquettes.}. Furthermore, by setting the lattice constant to be $1$, the coordinate of a $d$-cube in a $D$-dimensional cubic lattice always contains $(D-d)$ integers and $d$ half-integers. For example, in a $3$-dimensional cubic lattice,   $(0,0,0)$ refers to a $0$-cube, i.e., a point $(0,0,0)$; $(\frac{1}{2},0,0)$ refers to a $1$-cube, i.e., a link whose center is $(\frac{1}{2},0,0)$; $(\frac{1}{2},\frac{1}{2},0)$ refers to a $2$-cube, i.e., a plaquette whose center is $(\frac{1}{2},\frac{1}{2},0)$;  $(\frac{1}{2},\frac{1}{2},\frac{1}{2})$ refers to a $3$-cube, i.e., a common cube whose center is $(\frac{1}{2},\frac{1}{2},\frac{1}{2})$.

Next, we explain the meaning of ``nearest to''. In a $D$-dimensional cubic lattice, for $\gamma_{d_i}=(x_1,x_2,\cdots,x_D)$ and $\gamma_{d_j}=(y_1,y_2,\cdots,y_D)$, we say they are nearest to each other if and only if:
\begin{align*}
L_1(\gamma_{d_i},\gamma_{d_j})\equiv& |x_1-y_1|+|x_2-y_2|+\cdots+|x_D-y_D|\\=&\frac{|d_i-d_j|}{2},\ (d_i\neq d_j),\\
L_1(\gamma_{d_i},\gamma_{d_j})\equiv& |x_1-y_1|+|x_2-y_2|+\cdots+|x_D-y_D|\\=&1,\ (d_i=d_j).
\end{align*}
Moreover, given $d_i<d_j$, $\gamma_{d_i}$ being inside $\gamma_{d_j}$ is equivalent to that $\gamma_{d_i}$ being nearest to $\gamma_{d_j}$. A more detailed introduction of this notation is given in Sec.~II of Ref.~\cite{ye19a}.
 
\begin{figure*}[t]
	\includegraphics[width=0.53\textwidth]{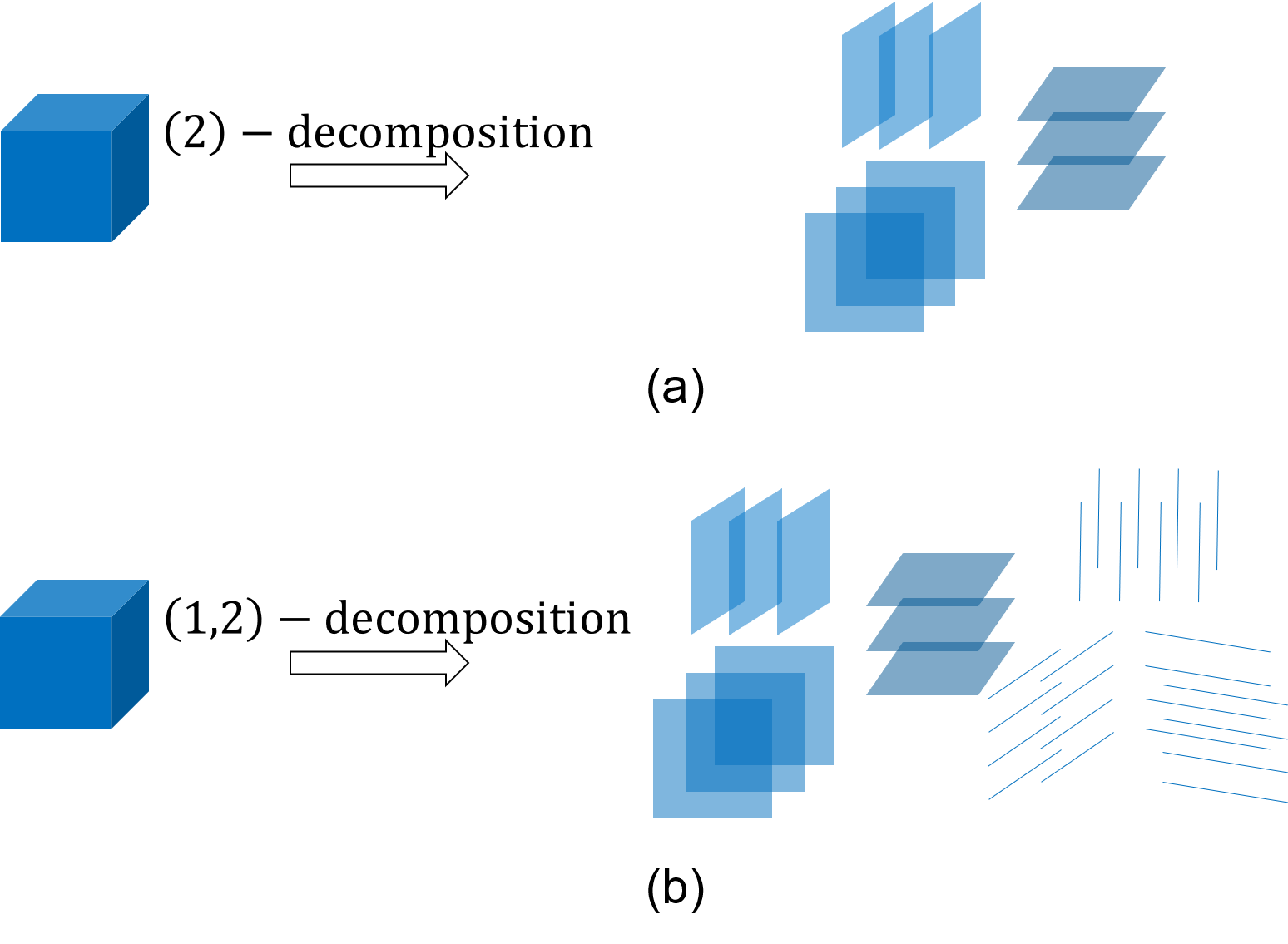}
	\caption{\textbf{Decompositions of 3-torus for X-cube model, where spins are located on links.} As we can see, while the decomposition in (a) shows a regular $2$-dimensional foliation of the original $3$-manifold, the decomposition in (b) additionally gives a foliation structure to each $2$-dimensional leaf space in (a).}
	\label{fig_space_decom}
\end{figure*}

Furthermore, in order to give an intuitive picture of the definition of $[d_n,d_s,d_l,D]$ models, here we give a heuristic introduction of foliation. For a more detailed discussion, see Ref.~\cite{hardorp1980all}. 
In general, foliation is the partition of a manifold into a set of submanifolds, where each  submanifold is  dubbed ``leaf''. For a regular foliation, where all leaves are of the same dimension, the dimension of leaves is defined as the dimension of the foliation. In this paper, as we only consider lattices defined on toric manifolds and regular foliations, we will treat leaves simply as sublattices. An example of foliation is pictorially illustrated in Fig.~\ref{fig_space_decom}.
For foliated fracton orders, two systems belong to the same phase if they can be transformed to each other under addition or removal of subdimensional topologically ordered phases. Therefore, their physical properties, like ground state degeneracy, have direct relationship with their foliation structures. A large amount of type-I fracton orders have been proved to be foliated fracton orders, like X-cube model and checkerboard model  \cite{Shirley2019,PhysRevB.99.115123,Ma2017}.
 
Last but not least, inspired by the idea of higher-order topological insulator  \cite{2018SciA....4..346S}, here we give a rough definition of higher-order boundary in cubic and hypercubic lattices, in order to refer to the higher-dimensional analogs of hinges of a cuboid. 
Firstly, we consider the $d$-dimensional analogs of cuboid, denoted as $S^{d}$. For example, an $S^1$ is a straight string, an $S^2$ is a flat membrane, and an $S^3$ is a cuboid\footnote{Here it is important to notice that $S^d$ is always ``flat''. For example, an ``L''-shaped string is composed of $2$ $S^1$'s, and the turning point of the string should be recognized as a vertex emanating $2$ perpendicular $S^1$'s}. See Ref.~\cite{ye19a} for more details of $S^n$ notations.
Then, for such an $S^{d}$, we can define its $d_i$-dimensional boundary as follows (here $d_i<d$ is assumed):

\begin{itemize}
	\item The $d_i$-dimensional boundary of an $S^d$ is composed of $\gamma_{d_i}$'s.
	\item A $\gamma_{d_i}$ belongs to the $d_i$-dimensional boundary of the $S^d$ if and only if it is nearest to exactly one $\gamma_d$ in the $S^d$.
\end{itemize}

We can check that such a definition is consistent with our intuition. For example, the $0$-dimensional boundary of a plaquette is composed of the $4$ vertices of the plaquette, and the $1$-dimensional boundary of a common cube is composed of the $12$ hinges of the cube. While for a cuboid extended along one spatial direction, we can see that its $1$-dimensional boundary is composed of only $4$ hinges. 

\subsection{Review of $[d_n,d_s,d_l,D]$ models}
\label{subsec:review}
As we shall exemplify our GSD computing algorithm with the $[d_n,d_s,d_l,D]$ models proposed in Ref.~\cite{ye19a},   it would be beneficial to give a review of the definition of the models at first. 
$[d_n,d_s,d_l,D]$ models are defined on $D$-dimensional cubic lattice with $1/2$-spins as basic degrees of freedom. There is a spin (i.e., a qubit)  at the center of each  $\gamma_{d_s}$. The Hamiltonian is defined as below:

\begin{align}
\label{eq:the_Hamiltonian}
H_{[d_n,d_s,d_l,D]} = -J\sum_{\{{\gamma}_{D}\}} A_{{\gamma}_{D}} - K \sum_{\{{\gamma}_{d_n}\}} \sum_{l} B^l_{{\gamma}_{d_n}},
\end{align}
where $J,K>0$ are coupling constants. $d_n<d_s<d_l<D$ is assumed. Apparently, there are two typical terms in the Hamiltonian, namely, $A$-term and $B$-term. 
\begin{itemize}
\item First,  given a $\gamma_{D}$, the Hermitian operator $A_{{\gamma}_{D}}$   is the product of all $x$-components (i.e., $\sigma^x$) of spins whose corresponding $\gamma_{d_s}$'s   are nearest to the $\gamma_D$.  The definition of ``nearest to'' can be found in Sec.~\ref{subsec_geo_not}.

\item Second, in order to determine the $B^l_{{\gamma}_{d_n}}$ term,   we need to choose a $d_l$-dimensional sublattice $l$ in which the $\gamma_{d_n}$ cube is embedded\footnote{By ``sublattice'', we mean that the sublattice is a part of the whole $D$-dimensional cubic lattice in the usual sense. For example, a flat plane formed by plaquettes is a $d_l$=2 sublattice.  By ``embed'', we mean that the $\gamma_{d_n}$ cube is exactly an ingredient of the sublattice.}. Then we need to find all $\gamma_{d_s}$'s which are not only nearest to the ${\gamma}_{d_n}$ but also totally embedded in $l$. Then   $B^l_{{\gamma}_{d_n}}$ is just the product of the $z$-components of all the spins at the centers of such $d_s$-cubes. 
\end{itemize}
 
 We note that, in general, there are infinite ``parallel'' sublattices if we only specify  $d_l$ orthogonal directions.   But the requirement of ``$\gamma_{d_n}$ being embedded in the sublattice'' unambiguously leads to a  unique sublattice.  Therefore, we only need to give $d_l$ orthogonal directions that unambiguously constitute the label $l$. 
 For example, in X-cube model, where $d_l=2$, we use $\langle \hat{x}_1 \hat{x}_2 \rangle$, $\langle \hat{x}_1 \hat{x}_3 \rangle$ and $\langle \hat{x}_2 \hat{x}_3 \rangle$ as the superscripts of $B$ terms. See Eq.~(\ref{eq:x-cube_B_term}) for an instance.

According to the definition of   Hamiltonians, we can see that the numbers in the label of a $[d_n,d_s,d_l,D]$ model are just dimension indices. In Ref.~\cite{ye19a}, it has been proved that a $[d_n,d_s,d_l,D]$ model is exactly solvable when $d_n,d_s,d_l$ and $D$ satisfy$\tbinom{d_l-d_n}{d_s-d_n} \; \text{mod} \; 2=0$ together with $d_n<d_s<d_l<D$.
Among them, there are two specially interesting branches of such exactly solvable models, $[0,1,2,D]$ and $[D-3,D-2,D-1,D]$. In this paper, we will mainly focus on these two branches of models.

In $[D-3,D-2,D-1,D]$ models with $D \geq 4$, our previous work  \cite{ye19a} demonstrates that there are extended excitations with fracton physics. That is to say, both mobility and deformability of spatially extended excitations (e.g. loop excitations and membrane excitations) in these models can be restricted. What is more, some topological excitations can even have non-manifold shapes. Such excitations with non-manifold-like shapes are dubbed ``complex excitations''. For complex excitations, as long as the connectivity of their shapes is preserved, they cannot be deformed to be manifold-like objects by any local unitary operators. Similarly, we can also define (intrinsically) disconnected excitations, whose shapes cannot be deformed to be connected by any local unitary operators. In conclusion, topological excitations in $[d_n,d_s,d_l,D]$ models can be classified into $4$ sectors, which are respectively trivial excitations, simple excitations, disconnected excitations and complex excitations. The latter $2$ sectors only exist in fracton orders.

Following the aforementioned rules of construction of Hamiltonians, we express explicitly Hamiltonians of some models when $D=3,4$:
\begin{itemize} 
	\item \textbf{X-cube  model (i.e., $[0,1,2,3]$ model)} In our notation, X-cube model is denoted as $[0,1,2,3]$. Thus, $d_n=0$, $d_s=1$, $d_l=2$ and $D=3$. Since $d_s=1$, spins are located at the centers of links. The Hamiltonian is composed of $B^l_{{\gamma}_{0}}$ and $A_{{\gamma}_{3}}$ terms defined on vertices and cubes respectively. And since we have $d_l=2$, such a $B^l_{{\gamma}_{0}}$ term equals to the product of $4$ $\sigma^z$'s sitting on the $4$ links which are (a) embedded in the plane $l$ and (b) nearest to the vertex ${\gamma}_{0}$. Similarly, an $A_{{\gamma}_{3}}$ term is the product of $12$ $\sigma^x$'s sitting on the $12$ links that are nearest to the $\gamma_3$. For example, for a common cubic lattice, we have
	\begin{align}
	\label{eq:x-cube_B_term}
	B^{\langle\hat{x}_1,\hat{x}_2 \rangle}_{(0,0,0)}=\sigma^z_{(0,\frac{1}{2},0)} \sigma^z_{(-\frac{1}{2},0,0)} \sigma^z_{(\frac{1}{2},0,0)} \sigma^z_{(0,-\frac{1}{2},0)},
	\end{align}  
	and
	\begin{align}
	\begin{split} A_{(\frac{1}{2},\frac{1}{2},\frac{1}{2})}=&\sigma^x_{(0,0,\frac{1}{2})} \sigma^x_{(0,1,\frac{1}{2})} \sigma^x_{(1,0,\frac{1}{2})} \sigma^x_{(1,1,\frac{1}{2})}\\ &\sigma^x_{(0,\frac{1}{2},0)} \sigma^x_{(0,\frac{1}{2},1)} \sigma^x_{(1,\frac{1}{2},0)} \sigma^x_{(1,\frac{1}{2},1)}\\ &\sigma^x_{(\frac{1}{2},0,0)} \sigma^x_{(\frac{1}{2},0,1)} \sigma^x_{(\frac{1}{2},1,0)} \sigma^x_{(\frac{1}{2},1,1)},
	\end{split}
	\end{align}
	where subscript coordinates like $(0,0,0)$ and $(1,0,\frac{1}{2})$ respectively refer to a $\gamma_0$ and a $\gamma_1$, as they are exactly the geometric centers of these two objects. In X-cube model, we have fractons and lineons as fundamental excitations. These excitations respectively correspond to the eigenvalue flips of $A_{{\gamma}_{3}}$ and $B^l_{{\gamma}_{0}}$ terms.
	
	Here we also give a brief introduction of the ground states of X-cube model (for a more detailed review of X-cube model, see Ref.~\cite{2020Fracton,Nandkishore2019}). As a stabilizer code model, a ground state of X-cube model $|\phi\rangle$ should satisfy the following conditions:
	\begin{align*}
	B^l_{\gamma_0} |\phi \ket = |\phi \ket,\ \forall \gamma_0,l\,;\,A_{\gamma_3} |\phi \ket = |\phi \ket,\ \forall \gamma_3\,.
	\end{align*}
	Hence, in $\sigma^z$ basis, such a ground state $|\phi \ket$ must be an equal weight superposition of $S^1$'s (i.e. straight strings) with vertices emanating $3$ perpendicular $S^1$'s. Similar to toric code model, independent ground states of X-cube model can be distinguished by the action of non-local logical operator $W(S^1)=\prod_{\gamma_1 \in S^1} \sigma^x_{\gamma_1}$ with non-contractible $S^1$. Besides, in X-cube model, four $S^1$'s that compose the $1$-dimensional boundary of an $S^3$ extended in one direction can be created or annihilated by applying $A_{\gamma_3}$ stabilizers. From another perspective, it means that one single non-local string can be ``split'' to $3$ different non-local strings under the action of $A_{\gamma_3}$ stabilizers. 
	
	\item \textbf{$[0,1,2,4]$ model} The Hamiltonian of $[0,1,2,4]$ model is composed  of $B^l_{{\gamma}_{0}}$ and $A_{{\gamma}_{4}}$ terms. And since we have $d_l=2$, such a $B^l_{{\gamma}_{0}}$ term equals to the product of $4$ $\sigma^z$'s sitting on the $4$ links which are (a) embedded in the plane $l$ and (b) nearest to the vertex ${\gamma}_{0}$. Similarly, an $A_{{\gamma}_{4}}$ term is the product of $32$ $\sigma^x$'s sitting on the $32$ links that are nearest to the $\gamma_4$. For a common hypercubic lattice, we have
	\begin{align}
	B^{\langle\hat{x}_1,\hat{x}_2 \rangle}_{(0,0,0,0)}=\sigma^z_{(0,\frac{1}{2},0,,0)} \sigma^z_{(-\frac{1}{2},0,0,0)} \sigma^z_{(\frac{1}{2},0,0,0)} \sigma^z_{(0,-\frac{1}{2},0,0)},
	\end{align}  
	and
	\begin{align}
	\begin{split}
	A_{(\frac{1}{2},\frac{1}{2},\frac{1}{2},\frac{1}{2})}=&\sigma^x_{(0,0,\frac{1}{2},0)} \sigma^x_{(0,1,\frac{1}{2},0)} \sigma^x_{(1,0,\frac{1}{2},0)} \sigma^x_{(1,1,\frac{1}{2},0)}\\ &\sigma^x_{(0,\frac{1}{2},0,0)} \sigma^x_{(0,\frac{1}{2},1,0)} \sigma^x_{(1,\frac{1}{2},0,0)} \sigma^x_{(1,\frac{1}{2},1,0)}\\ &\sigma^x_{(\frac{1}{2},0,0,0)} \sigma^x_{(\frac{1}{2},0,1,0)} \sigma^x_{(\frac{1}{2},1,0,0)} \sigma^x_{(\frac{1}{2},1,1,0)}\\
	&\sigma^x_{(0,0,\frac{1}{2},1)} \sigma^x_{(0,1,\frac{1}{2},1)} \sigma^x_{(1,0,\frac{1}{2},1)} \sigma^x_{(1,1,\frac{1}{2},1)}\\ &\sigma^x_{(0,\frac{1}{2},0,1)} \sigma^x_{(0,\frac{1}{2},1,1)} \sigma^x_{(1,\frac{1}{2},0,1)} \sigma^x_{(1,\frac{1}{2},1,1)}\\ &\sigma^x_{(\frac{1}{2},0,0,1)} \sigma^x_{(\frac{1}{2},0,1,1)} \sigma^x_{(\frac{1}{2},1,0,1)} \sigma^x_{(\frac{1}{2},1,1,1)}\\
	&\sigma^x_{(0,0,0,\frac{1}{2})}
	\sigma^x_{(0,0,1,\frac{1}{2})}
	\sigma^x_{(0,1,0,\frac{1}{2})}
	\sigma^x_{(0,1,1,\frac{1}{2})}\\
	&\sigma^x_{(1,0,0,\frac{1}{2})}
	\sigma^x_{(1,0,1,\frac{1}{2})}
	\sigma^x_{(1,1,0,\frac{1}{2})}
	\sigma^x_{(1,1,1,\frac{1}{2})}.\\
	\end{split}
	\end{align} 
	In $[0,1,2,4]$ model, we have fractons and lineons as fundamental excitations. These excitations respectively correspond to the eigenvalue flips of $A_{{\gamma}_{4}}$ and $B^l_{{\gamma}_{0}}$ terms. A pictorial demonstration of such an $A_{\gamma_4}$ term is given in Fig.~\ref{fig_4cube_term}.
	
	\item \textbf{$[1,2,3,4]$ model} The Hamiltonian of $[1,2,3,4]$ model is composed of $B^l_{{\gamma}_{1}}$ and $A_{{\gamma}_{4}}$ terms. And since we have $d_l=3$, such a $B^l_{{\gamma}_{1}}$ term equals to the product of $4$ $\sigma^z$'s sitting on the $4$ plaquettes which are (a) embedded in the 3-dimensional space $l$ and (b) nearest to the link ${\gamma}_{1}$. Similarly, an $A_{{\gamma}_{4}}$ term is the product of $24$ $\sigma^x$'s sitting on the $24$ plaquettes that are nearest to the $\gamma_4$. For a common hypercubic lattice, we have
	\begin{align}
	\label{eq:B terms}
	\begin{split}
	&B^{\langle \hat{x}_1,\hat{x}_2,\hat{x}_4\rangle}_{(0,0,0,\frac{1}{2})} = \sigma^z_{(\frac{1}{2},0,0,\frac{1}{2})} \sigma^z_{(-\frac{1}{2},0,0,\frac{1}{2})} \sigma^z_{(0,\frac{1}{2},0,\frac{1}{2})} \sigma^z_{(0,-\frac{1}{2},0,\frac{1}{2})},\\
	\end{split}
	\end{align} 
	and
	\begin{align}
	\label{eq:A terms}
	\begin{split}
	A_{(\frac{1}{2},\frac{1}{2},\frac{1}{2},\frac{1}{2})} =& \sigma^x_{(0,0,\frac{1}{2},\frac{1}{2})} \sigma^x_{(0,1,\frac{1}{2},\frac{1}{2})}  \sigma^x_{(1,0,\frac{1}{2},\frac{1}{2})}  \sigma^x_{(1,1,\frac{1}{2},\frac{1}{2})}\\ &\sigma^x_{(0,\frac{1}{2},0,\frac{1}{2})}  \sigma^x_{(0,\frac{1}{2},1,\frac{1}{2})} 
	\sigma^x_{(1,\frac{1}{2},0,\frac{1}{2})} \sigma^x_{(1,\frac{1}{2},1,\frac{1}{2})}\\    &\sigma^x_{(0,\frac{1}{2},\frac{1}{2},0)} \sigma^x_{(0,\frac{1}{2},\frac{1}{2},1)}  \sigma^x_{(1,\frac{1}{2},\frac{1}{2},0)} \sigma^x_{(1,\frac{1}{2},\frac{1}{2},1)}\\  &\sigma^x_{(\frac{1}{2},0,0,\frac{1}{2})}  \sigma^x_{(\frac{1}{2},0,1,\frac{1}{2})}  \sigma^x_{(\frac{1}{2},1,0,\frac{1}{2})} \sigma^x_{(\frac{1}{2},1,1,\frac{1}{2})}\\  &\sigma^x_{(\frac{1}{2},0,\frac{1}{2},0)} \sigma^x_{(\frac{1}{2},0,\frac{1}{2},1)} \sigma^x_{(\frac{1}{2},1,\frac{1}{2},0)}  \sigma^x_{(\frac{1}{2},1,\frac{1}{2},1)}\\  &\sigma^x_{(\frac{1}{2},\frac{1}{2},0,0)} \sigma^x_{(\frac{1}{2},\frac{1}{2},0,1)}  \sigma^x_{(\frac{1}{2},\frac{1}{2},1,0)} \sigma^x_{(\frac{1}{2},\frac{1}{2},1,1)}.\\
	\end{split}
	\end{align}

In $[1,2,3,4]$ model, we have fractons and $(1,2)$-type excitations as fundamental excitations. These excitations respectively correspond to the eigenvalue flips of $A_{{\gamma}_{4}}$ and $B^l_{{\gamma}_{1}}$ terms. Here $(1,2)$-type means that the excitation is intrinsically $1$-dimensional (i.e. it is a string), and its mobility and deformability are restricted in a $2$-dimensional subspace. For more details of $(m,n)$-type excitations, see Ref.~\cite{ye19a}.

What is more, in $[1,2,3,4]$ model, there are also complex excitations \textit{chairon} and \textit{yuon}, whose shapes are non-manifold-like. As an example, the shape of a yuon is composed of $3$ strings which share one pair of endpoints. Due the the deformability restriction, such a non-manifold shape of a yuon is invariant under local unitary transformations, as long as its connectivity is preserved.

\begin{figure}[t]
	\includegraphics[width=0.45\textwidth]{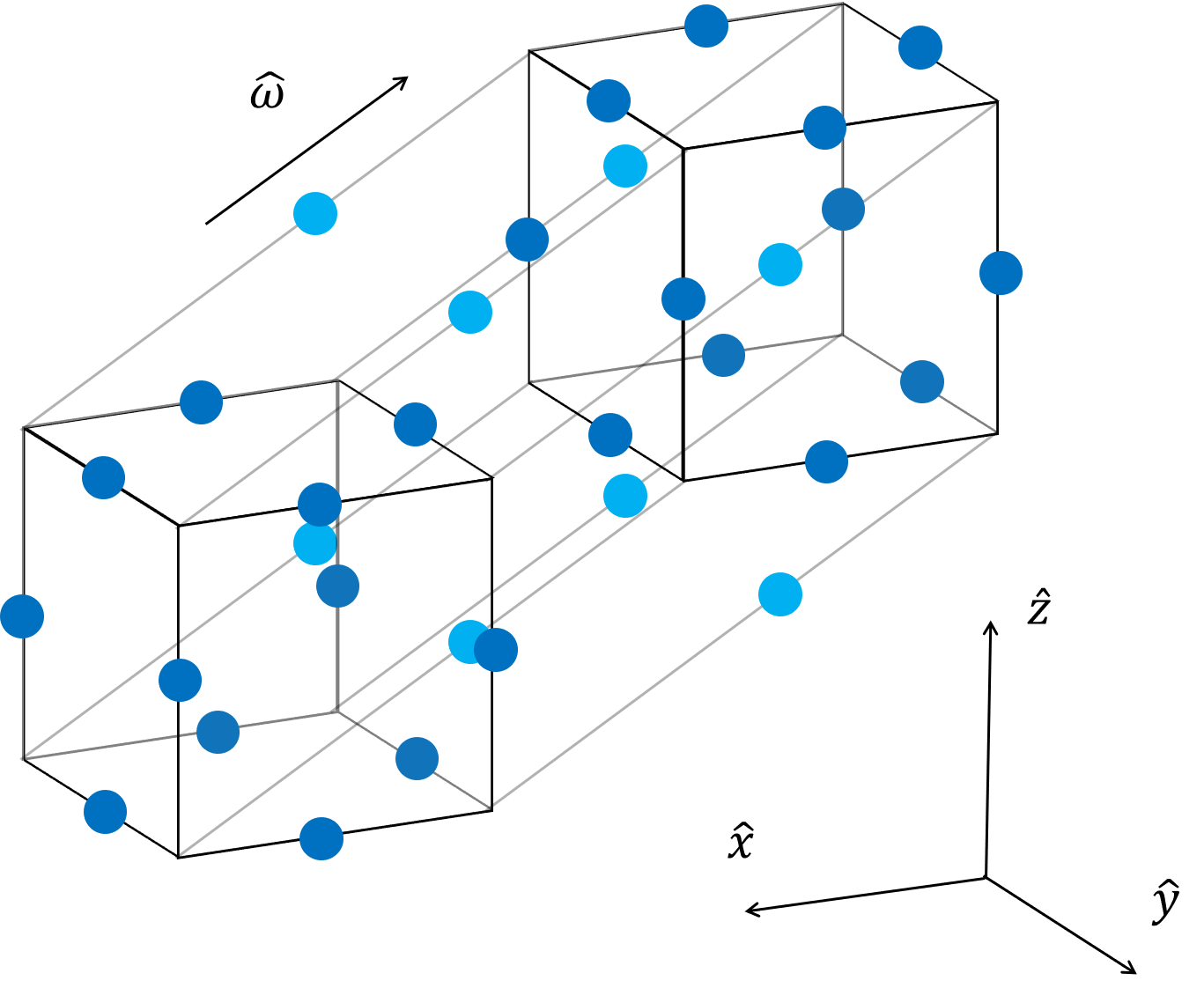}
	\caption{\textbf{Pictorial demonstration of an $A_{\gamma_4}$ term in $[0,1,2,4]$ model on a $4$-dimensional lattice.} Spins are located on links, and they are represented by blue dots. While for clarity, spins along the fourth spatial dimension denoted as $\hat{\omega}$ are highlighted with light blue. As we can see, the $A_{\gamma_4}$ is the product of $32$ $\sigma^x$ operators on the $32$ links of a $4$-dimensional hypercube.}
	\label{fig_4cube_term}
\end{figure} 

\end{itemize}

\section{Representing ground states with lower-dimensional data}
\label{subsec_res_of_gs}

As we are trying to compute the GSD of a $[d_n,d_s,d_l,D]$ model by data from lower-dimensional subsystems\footnote{In this paper, we use ``sublattice'' and ``subsystem'' alternatively.}, it is important to define the subsystem structures strictly. So in this section, we will introduce the definition of decomposition and restriction of lattices, Ising configurations, and ground states. And finally, we prove that in certain cases, there is a one-to-one correspondence between a collection of  subsystem ground state sectors ($\mathsf{SGS}$) satisfying certain \textit{consistent conditions} (to be defined in this section) and a ground state of the original system. Consequently, it is plausible to compute the GSD of a $[d_n,d_s,d_l,D]$ model by counting the number of possible combinations of subsystem states.

\subsection{Decomposition of lattices}

To define a decomposition of a ground state in a general $[d_n,d_s,d_l,D]$ model, we find that it is necessary to first  consider a decomposition of manifolds. More concretely, because the ``ground state degeneracy'' of a leaf can also be subextensive, in some higher-dimensional models, we can consider ``leaves of a leaf.'' From another view, a higher-dimensional $[d_n,d_s,d_l,D]$ model may admit foliation structures of various dimensions. As a result, we believe it is beneficial to define a ``decomposition'' of the base manifold. Besides, in this paper, as we are mainly interested in lattice models, manifolds are always assumed to be equipped with a lattice structure (i.e., cellulation) given by the definition of a $[d_n,d_s,d_l,D]$ model. \textit{Therefore, when the lattice structure is specified, we will alternatively use ``(sub-)lattice'' and ``(sub-)manifold''. For simplicity, in the paper we only consider toric base manifolds.}

For a base manifold $M^D$, we define a $(d_1,d_2,\cdots,d_k)$-decomposition (denoted by $\mathcal{M}$) as a collection composed of sets of lower-dimensional manifolds 
\begin{align}
\mathcal{M}\equiv\{M^{d_1}\}\cup\{ M^{d_2}\}\cup\cdots\cup\{M^{d_k}\},\label{equation_decomp_definition}
\end{align}
where the notation $\{M^{d_i}\}$ denotes a set in which all manifolds are $d_i$-dimensional. The above definition of decomposition, i.e.,  Eq.~(\ref{equation_decomp_definition}) satisfies the following (without loss of generality, we always assume $d_1<d_2<\cdots <d_k$):
\begin{itemize}
	\item For any $\{M^{d_i}\}$, we have $\cup_{M\in\{M^{d_i}\}} M = M^D$. Specially, for lattices, this condition requires that every spin   belongs to at least one sublattice $M\in\{M^{d_i}\},\forall \ d_i$.
	\item For any two sets of lower-dimensional manifolds $\{M^{d_i}\}$ and $\{M^{d_j}\}$ with $d_i<d_j$, we require that for any $M^{d_j}\in\{M^{d_j}\}$, there exists a subset $\{\tilde{M}^{d_i}\}\subset \{M^{d_i}\}$, such that $\cup_{M^{d_i}\in\{\tilde{M}^{d_i}\}} M^{d_i} = M^{d_j}$. For lattices, this condition requires that every spin in $M^{d_j}$ belongs to at least one sublattice $M^{d_i}\in\{\tilde{M}^{d_i}\}$.
\end{itemize}

Here, it should be noted that these conditions show that a decomposition of a manifold equipped with a lattice structure can only be determined when the $[d_n,d_s,d_l,D]$ model is specified. The reason is rather obvious: specifying a model can tell how spins (qubits) are distributed spatially. In the simplest decomposition,  the collection $\mathcal{M}$ contains only subsystems  of the same dimension, i.e.,   $\mathcal{M}=\{M^{d_i}\}$ with $d_i<D$,   a regular foliation of $M^D$ is achieved \cite{hardorp1980all}. When the collection contains sets of sublattices of different dimensions, the decomposition gives foliation structures with different dimensional sublattices, too (see Fig.~\ref{fig_space_decom}).

In this paper, as we do not discuss about the influence of the topology of sublattices in detail, we always assume that the decompositions are isotropic and composed of toric sublattices.
The effect of different boundary conditions is definitely very interesting, which will be  studied separately.

\subsection{Decomposition of Ising configurations and quantum operators}

\begin{figure}[t]
	\includegraphics[width=0.47\textwidth]{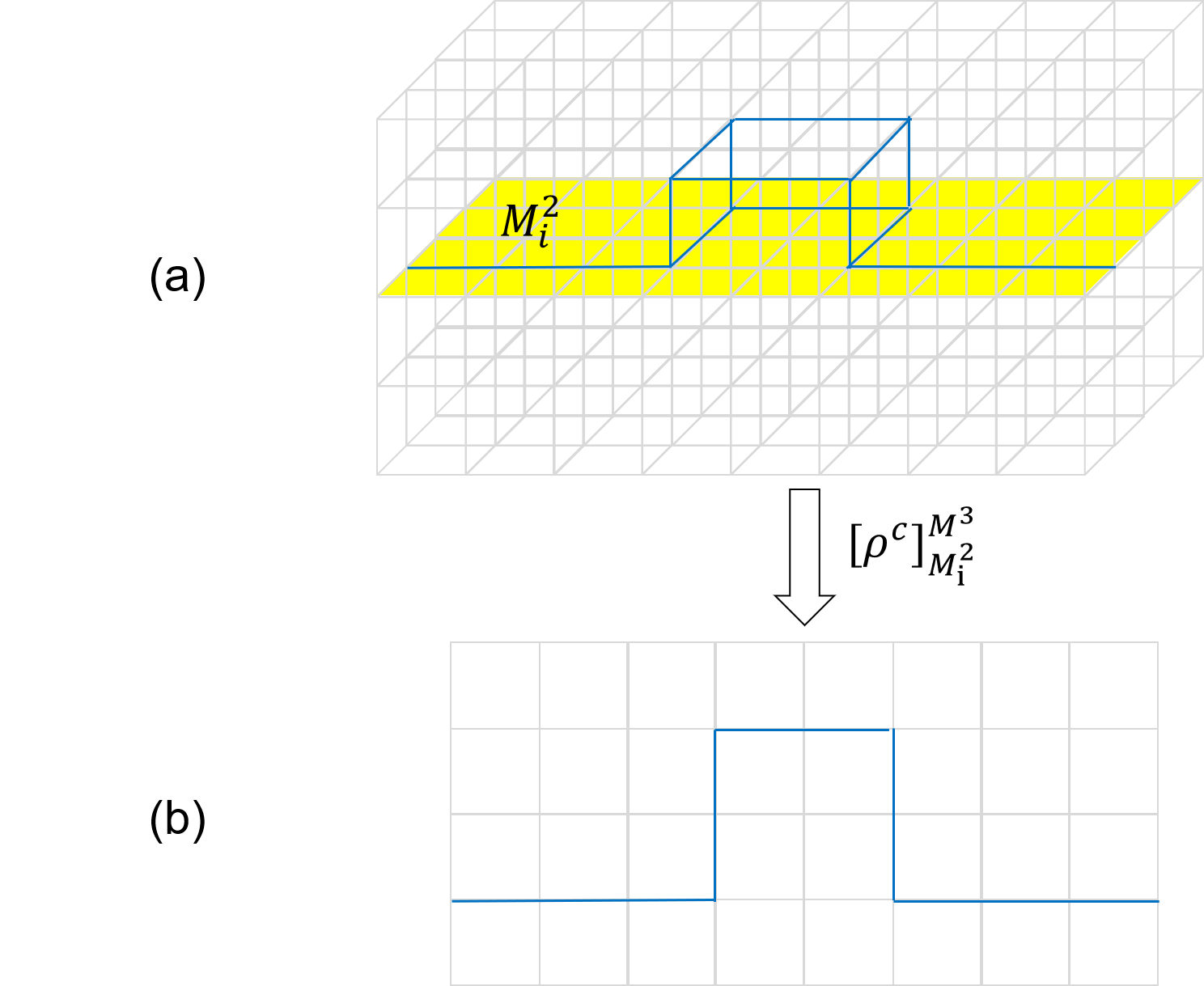}
	\caption{\textbf{Restriction of Ising configurations in X-cube model.}  Spins in X-cube model  are located at centers of links. In (a), links with down spins   are marked in blue. $M^2_i$ plane  is highlighted in yellow. In (b),  the restriction of the blue string composed of flipped spins onto subsystem $M^2_i$ is described. The restriction leads to a ``humplike'' shape Ising configuration.}
	\label{fig_config_res}
\end{figure}

Next, for a given decomposition $\mathcal{M}$ of the base manifold $M^D$, we can define the decomposition of Ising configurations and operators.   Without special note, in this paper we only consider $\frac{1}{2}$-spins as basic degrees of freedom, and the $\sigma^z$ (Ising) basis is always assumed. 
Given a model defined on $M^D$ and one of its decomposition $\mathcal{M}$, when there is an inclusion $I:M^{d_1}_i\hookrightarrow M^{d_2}_j$, we can define the restriction of an arbitrary Ising configuration: 
\begin{align} 
[\rho^{\mathrm{c}}]^{M^{d_2}_j}_{M^{d_1}_i}(c_{M^{d_2}_j})=c_{M^{d_1}_i}\,.
\end{align}
It maps $c_{M^{d_2}_j}$ to $c_{M^{d_1}_i}$ by simply dropping all degrees of freedom out of $M^{d_1}_i$. Here, $c_{M^{d_1}_i}$ and $c_{M^{d_2}_j}$ are Ising configurations respectively defined on $M^{d_1}_i$ and $M^{d_2}_j$. In the notation $\rho^{\mathrm{c}}$, the superscript $\mathrm{c}$ literally stands for ``configuration''. Then, for a given Ising configuration $c$ and manifold decomposition $\mathcal{M}$, we can define the decomposition of the Ising configuration $c$ on $\mathcal{M}$ as $\Omega^{\mathrm{c}}_{\mathcal{M}}(c)$, by taking the set of all restrictions of $c$ onto sublattices inside $\mathcal{M}$. That is to say, we have
\begin{align} \Omega^{\mathrm{c}}_{\mathcal{M}}(c)\equiv \{c_{M^{d_1}}\}\cup\{ c_{M^{d_2}}\}\cup\cdots\cup\{c_{M^{d_k}}\},
\end{align}
where $c_{M^{d_i}}$ refers to the restriction of the Ising configuration $c$ onto a $d_i$-dimensional sublattice $M^{d_i}$. According to the definition of $\mathcal{M}$, the original Ising configuration $c$ can be reconstructed from one of its decomposition $\Omega^{\mathrm{c}}_{\mathcal{M}}(c)$. A pictorial demonstration of the restriction of Ising configurations is given in Fig.~\ref{fig_config_res}.

Similarly, with a given model and a decomposition $\mathcal{M}$, for a class of specially interesting operators, which are composed of $\sigma^x$ operators (called ``$X$-operator'' for simplicity), we can also define their restriction and decomposition. When there is an inclusion $I:M^{d_1}_i\hookrightarrow M^{d_2}_j$, the restriction 
\begin{align}
[\rho^{\mathrm{o}}]^{M^{d_2}_j}_{M^{d_1}_i}(o_{M^{d_2}_j})=o_{M^{d_1}_i}
\end{align}
maps $o_{M^{d_2}_j}$ to $o_{M^{d_1}_i}$ by dropping all $\sigma^x$ operators defined on $\gamma_{d_s}$'s out of $M^{d_1}_i$. Here, $o_{M^{d_2}_j}$ and $o_{M^{d_1}_i}$ are respectively $X$-operators totally supported on $M^{d_2}_j$ and $M^{d_1}_i$. In the notation $ \rho^{\mathrm{o}} $, the superscript $\mathrm{o}$ literally stands for ``operator''. Again, for a given $X$-operator $o$ and manifold decomposition $\mathcal{M}$, we can define the decomposition of the $X$-operator $o$ on $\mathcal{M}$ as $\Omega^{\mathrm{o}}_{\mathcal{M}}(o)$, by taking the set of all restrictions of $o$ onto sublattices inside $\mathcal{M}$. That is to say, we have
\begin{align}
\Omega^{\mathrm{o}}_{\mathcal{M}}(o)\equiv \{o_{M^{d_1}}\}\cup\{ o_{M^{d_2}}\}\cup\cdots\cup\{o_{M^{d_k}}\}, 
\end{align} 
where $o_{M^{d_i}}$ refers to the restriction of the $X$-operator $o$ onto a $d_i$-dimensional sublattice $M^{d_i}$. Since all $\sigma^x$ operators commute with each other, an arbitrary $X$-operator $o$ can also be reconstructed from one of its decomposition $\Omega^{\mathrm{o}}_{\mathcal{M}}(o)$.

\subsection{Subsystem ground state sector ($\mathsf{SGS}$) as a holographic shadow of ground states}
\label{subsubsec_decom_states}

Then, we can define the decomposition of a ground state sector with a given manifold decomposition $\mathcal{M}$\footnote{We use ``ground state sectors'' instead of ``ground states'' in this section to emphasize that they can be regarded as sets of  Ising  configurations, as demonstrated in this subsection.}. Nevertheless,  we have to at first consider how to define a ground state on a sublattice, as it is not as intuitive as the restriction of an  Ising  configuration. To solve this problem, we need to utilize the properties of ground states of $[d_n,d_s,d_l,D]$ models. For a ground state $|{x}\rangle$ of a $[d_n,d_s,d_l,D]$ model, it must satisfy the following conditions:
\begin{align}
B^l_{\gamma_{d_n}} |{x}\rangle = |{x}\rangle,\ \forall\ l,\gamma_{d_n}\,;  
A_{\gamma_{D}} |{x}\rangle = |{x}\rangle,\ \forall\ \gamma_{D}\,.
\end{align}
In $\sigma^z$ basis, the constraints given by $B^l_{\gamma_{d_n}}$ terms can be realized by requiring all Ising configurations in $|{x}\rangle$ to satisfy certain conditions. As an example, in X-cube model denoted as $[0,1,2,3]$, the  constraints given by $B^l_{\gamma_{0}}$ terms require that a configuration $c$ can only contain strings (formed by down spins) with trivalent vertices (see Sec.~\ref{subsec:review}). The constraints given by $A_{\gamma_{D}}$ terms require that, all  Ising  configurations, which can be transformed to each other by applying $A_{\gamma_{D}}$ operators,   must be equally superpositioned in a ground state $|{x}\rangle$.  \textit{Consequently, a ground state $|{x}\rangle$ of a $[d_n,d_s,d_l,D]$ model can be regarded as a set of  Ising configuration $c$'s, and the set satisfies constraints given by both $A$ and $B$ terms.}

Similarly, we can define a \textit{subsystem ground state sector} ($\mathsf{SGS}$) on sublattice $M^{d_i}$, denoted as $x_{M^{d_i}}$. The SGS is a set of $c_{M^{d_i}}$'s. This set satisfies the following conditions: 
\begin{itemize}
	\item   $c_{M^{d_i}}$   can be obtained as the restriction of a configuration $c$ satisfying $B$ constraints onto $M^{d_i}$. This condition results from the $B$ constraints over original ground states.
	\item Two $c_{M^{d_i}}$'s belong to the same $\mathsf{SGS}$, i.e., $x_{M^{d_i}}$, if and only if they can be connected under the action of $[\rho^{\mathrm{o}}]^{M^D}_{M^{d_i}}(A_{\gamma_D})$. This condition results from the $A$ constraints over original ground states. In other words, all elements in the $\mathsf{SGS}$ denoted as ``$x_{M^{d_i}}$'' can be mapped to each other by $[\rho^{\mathrm{o}}]^{M^D}_{M^{d_i}}(A_{\gamma_D})$.  
\end{itemize}

Now we can see that, according to our definition of $\mathsf{SGS}$, there is a well-defined restriction of a ground state $x$\footnote{To stress that $x$ is a set of  Ising  configurations here, we omit the bracket notation.}: 
\begin{align}
[\rho^{\mathrm{x}}]^{M^D}_{M^{d_1}_i}(x)=x_{M^{d_1}_i}\,
\end{align}
which maps a ground state to an $\mathsf{SGS}$ on $M^{d_1}_i$ (to see the existence of this map, we only need to notice that by definition, the restriction of $A_{\gamma_D}$ operators does not change any subsystem ground state sectors). Then we can define the decomposition of a ground state $x$ on $\mathcal{M}$ following exactly the same manner as  Ising configurations and $X$-operators, by taking the set of all restrictions of $x$ onto sublattices inside $\mathcal{M}$. So we have
\begin{align} \Omega^{\mathrm{x}}_{\mathcal{M}}(x)\equiv \{x_{M^{d_1}}\}\cup\{ x_{M^{d_2}}\}\cup\cdots\cup\{x_{M^{d_k}}\},
\end{align}
where $x_{M^{d_i}}$ refers to the restriction of the ground state $x$ on a $d_i$-dimensional sublattice $M^{d_i}$. \textit{Intuitively, $\mathsf{SGS}$ can be regarded as a shadow of original ground states.}  $\{x_{M^{d_i}}\}$ is the shadow.

Finally, we indicate that a ground state $x$ of a $[d_n,d_s,d_l,D]$ model can be completely reconstructed from a decomposition of $x$ on $\mathcal{M}$, as long as (a) $\mathcal{M}$ contains $(d_s+1)$-dimensional subsystems, (b) $d_s-d_n=1$ and (c) the base manifold $M$ is a torus (a proof and some discussion about the cases when $M$ has more complicated topology are given in Appendix~\ref{sec_proof}).  For this reason, $\mathsf{SGS}$ can be regarded as a \textit{holographic shadow} of original ground states.

In this paper, we only consider $[d_n,d_s,d_l,D]$ models defined on toric base manifolds with $d_s-d_n=1$, and we always assume that our manifold decompositions contain $(d_s+1)$-dimensional subsystems. Thereupon, we can reconstruct ground states from $\mathsf{SGS}$'s. Nevertheless, not all combinations correspond to original ground states. In fact, a combination of $\mathsf{SGS}$'s has to satisfy certain \textit{consistent conditions} to be corresponding to an original ground state. Obviously, all such consistent combinations are decompositions of original ground states, so we can use this property as a formal definition of the consistent conditions. More concrete description of consistent conditions depend on concrete models.  In Sec.~\ref{subsec_x_cube}, we give a concrete example of consistent conditions in X-cube model.

In summary, once we have found the $\mathsf{SGS}$'s of a manifold decomposition $\mathcal{M}$ and their consistent conditions, the ground state degeneracy can be obtained by counting the number of consistent combinations of $\mathsf{SGS}$'s. Nevertheless, it seems that there is no general method to derive the consistent conditions systematically.

\subsection{Application in $[0,1,2,3]$ model (X-cube model)}
\label{subsec_x_cube}
 
As an example, let us consider the ordinary $(2)$-decomposition (i.e. $2$-dimensional foliation) of X-cube model on a $3$-torus \cite{Shirley2018}. As we have reviewed in Sec.~\ref{subsec:review}, the $B$ constraints of Ising configurations require that all configuration $c$'s can only have trivalent vertices. On $2$-dimensional subsystems, this constraint requires a $c_{M^2}=[\rho^{\mathrm{c}}]^{M^3}_{M^{2}}(c)$ in an $\mathsf{SGS}$ to be a closed string configuration. And the restriction of $A_{\gamma_3}$ stabilizers on $2$-dimensional subsystems are just plaquette operators
\begin{align}A_{\gamma_2}\equiv[\rho^{\mathrm{o}}]^{M^3}_{M^2_i}(A_{\gamma_3}) = \sigma^x_{\gamma_1^1}\sigma^x_{\gamma_1^2}\sigma^x_{\gamma_1^3}\sigma^x_{\gamma_1^4}\,,
\end{align}
where $\gamma_1^1,\ \gamma_1^2,\ \gamma_1^3,\ \gamma_1^4$ are four links of a plaquette $\gamma_2$. 

\begin{figure*}[t]
	\includegraphics[width=1\textwidth]{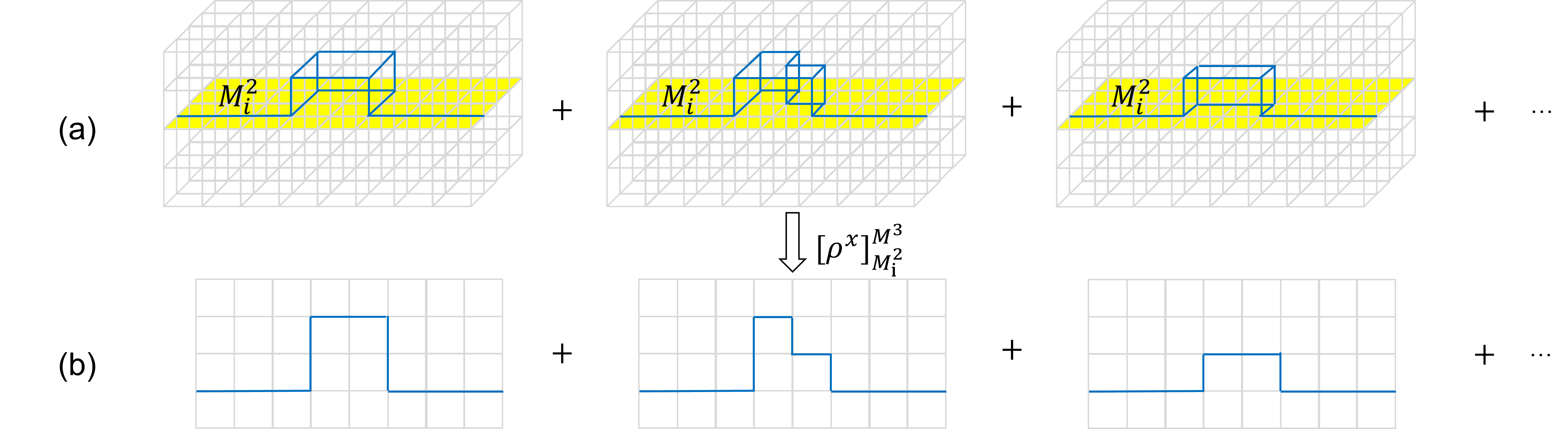}
	\caption{\textbf{Restriction of a ground state in X-cube model.} The notation in this figure is the same as in Fig.~\ref{fig_config_res}. As we can see, the restriction of the ground state of X-cube model in (a) is equivalent to a ground state of a 2D toric code model.}
	\label{fig_state_res}
\end{figure*}

Therefore, an X-cube ground state can be determined by a consistent combination of 2D $\mathsf{SGS}$'s, and here an $\mathsf{SGS}$ can be identified as a 2D toric code ground state. Fig.~\ref{fig_state_res} demonstrates such a restriction of a ground state. Then, we can see that the consistent condition is simply resulted from the fact that a logical operator $W(S^1) = \prod_{{\gamma_1}\in S^1} \sigma^x_{\gamma_1}$ in X-cube model simultaneously change the $\mathsf{SGS}$'s on $2$ intersecting $M^2$'s. In consequence, we can obtain all consistent combinations of $\mathsf{SGS}$'s by acting $W(S^1)$ operators on two intersecting $M^2$'s simultaneously from the combination where all $\mathsf{SGS}$'s have no non-local strings. 

In general, we develop a method to represent all such consistent combinations with coloring patterns of certain graphs. In Sec.~\ref{subsec_coloring}, we demonstrate this coloring method in detail, and give a concrete calculation of the GSD of X-cube model.

\section{General procedures and coloring method}
\label{subsec:overview}
 
\subsection{General   procedures}
 
Below we give a brief introduction of the key steps of the calculation.
For $[0,1,2,D]$ and $[D-3,D-2,D-1,D]$ models, similar to X-cube model (see Sec.~\ref{subsec_x_cube}), their $\mathsf{SGS}$'s on $M^{d_s+1}$'s can be identified as ground state sectors of some pure topological orders. That is to say, these $\mathsf{SGS}$'s can be obtained by exactly the same manner as ground states of some pure topological orders. As a result, in general, we can obtain the GSD of these models by the following steps:

\begin{itemize}
	\item First, determine all $\mathsf{SGS}$'s on $(d_s+1)$-dimensional subsystems $x_{M^{d_s+1}_i}=[\rho^{\mathrm{x}}]^{M^D}_{M^{d_s+1}_i}$ according to the definition of $\mathsf{SGS}$ given in Sec.~\ref{subsubsec_decom_states}.
	\item Second, find the consistent conditions between $\mathsf{SGS}$'s $x_{M^{d_s+1}_i}$'s.
	\item Third, calculate the number of consistent combinations of $\mathsf{SGS}$'s with a combinatorial method.
\end{itemize}

Here the counting of consistent combinations is not completely intuitive, since it involves the counting of high-dimensional objects. To avoid direct discussion about high dimensional geometric objects, it is convenient to use a combinatorial method, which converts the calculation to a coloring problem. In the rest part of this subsection, we will briefly demonstrate the construction of such coloring problems, and how to obtain GSD from them.

\subsection{Coloring method: general description and application in $[0,1,2,3]$ model (i.e., X-cube model)}
\label{subsec_coloring}

In $[D-3,D-2,D-1,D]$ and $[0,1,2,D]$ models with $\sigma^z$ basis, logical operators are all generated by $W(S^{d_s}) = \prod_{\gamma_{d_s} \in S^{d_s}} \sigma^x_{\gamma_{d_s}}$ operators, where $S^{d_s}$ is closed. Therefore, the $\mathsf{SGS}$'s on $M^{d_s+1}$'s are also distinguished by the action of $W(S^{d_s})$ operators (see Sec.~\ref{subsec:012D_models} and Sec.~\ref{subsec:-3-2-1D_models} for a more detailed demonstration). So the consistent condition of these $\mathsf{SGS}$'s is simply resulted from the geometric fact that an $S^{d_s}$ is always the intersection of $(D-d_s)$ $M^{d_s+1}$'s. Furthermore, as there are $\binom{D}{d_s}$ possible directions of $S^{d_s}$, and $W(S^{d_s})$ operators with $S^{d_s}$'s along different directions are independent, we can separate all ground states and $\mathsf{SGS}$'s into $\binom{D}{d_s}$ parts according to the direction of $S^{d_s}$. In general, the consistent condition of $x_{M^{d_s+1}_i}$'s in $[D-3,D-2,D-1,D]$ and $[0,1,2,D]$ models is that we can only change $\mathsf{SGS}$'s of the same part on $(D-d_s)$ intersecting $M^{d_s+1}$'s simultaneously.

Therefore, we can use a coloring problem to visualize the consistent combinations of $\mathsf{SGS}$'s. We can construct a \textit{characteristic graph} of a $[D-3,D-2,D-1,D]$ or $[0,1,2,D]$ model for a given part of ground states. The construction is composed of the following steps:
\begin{itemize}
	\item First, in a characteristic graph, we use a vertex to refer to an $S^{d_s}$ belonging to the given part (i.e. along certain direction), and a straight line composed of edges to refer to an $M^{d_s+1}$. 
	\item Second, lines intersecting at a vertex means that the corresponding $M^{d_s+1}$'s intersect at the $S^{d_s}$ represented by the vertex. 
	\item Then, we use colors of a line to refer to $\mathsf{SGS}$'s on the $M^{d_s+1}$. Because we only consider $\Z_2$ degrees of freedom in this paper, there are only $2$ colors of lines. So we can simply label lines as ``colored'' and ``uncolored''.\footnote{As all edges along the same line are always colored simultaneously, in this paper we can use ``edge'' and ``line'' interchangeably. This is different from the usual case of mathematical discussion about graphs.}
	\item Finally, we require that all coloring operations should be done for a group of lines intersecting at one vertex simultaneously. Besides, we only consider coloring patterns that can be obtained by coloring the completely uncolored pattern.
\end{itemize}

In this manner, we can use a coloring pattern to represent a consistent combination of $\mathsf{SGS}$'s, and the computation of GSD is just equivalent to counting such coloring patterns.

\begin{figure}[t]
	\includegraphics[width=0.45\textwidth]{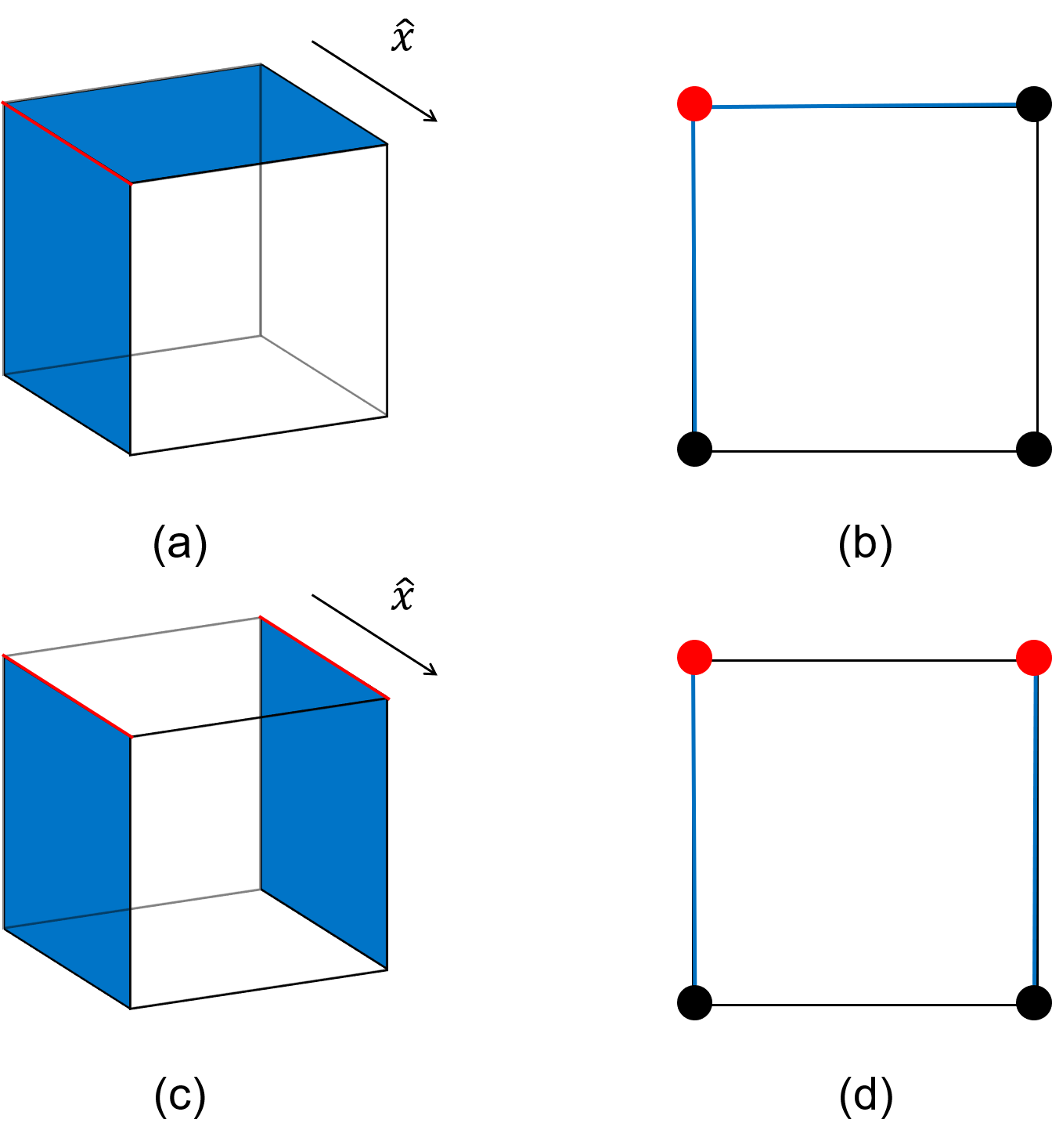}
	\caption{\textbf{An example of the correspondence between an Ising configuration with non-contractible loops in X-cube model and the associated characteristic graph.} The red bar in (a) refers to the action of a $W(S^1)$ operator, where $S^1$ is a non-contractible loop, and the leaves $M^2$ with odd number of $W(S^1)$ actions are highlighted with blue. In (b), the vertex corresponding to the $M^1$ with a $W(S^1)$ action is colored with red, and its associated lines are colored with blue. In (c), we add another non-contractible loop that is parallel to the first one. Then, since the top leaf has been acted twice, it restores to the uncolored state, and in its corresponding coloring pattern of characteristic map (d), the line represents the top leaf is also uncolored. }
	\label{fig:charac}
\end{figure}

As an example, we  consider X-cube model. A brief introduction of the ground states of X-cube model is given in Sec.~\ref{subsec:review}. In X-cube model, we have logical operators generated by $W(S^1) = \prod_{{\gamma_{d_s}}\in S^1} \sigma^x_{\gamma_{d_s}}$ operators. Thereupon, $\mathsf{SGS}$'s on $M^{2}$'s can also be changed by $W(S^1)$ operators. Obviously, $W(S^1)$ with $S^1$ along different directions are independent, so we can consider logical operators along the three spatial directions separately to obtain the GSD. Without loss of generality, here we consider the part of $S^1$'s along $\hat{x}_1$ direction. After that, we can see the consistent condition is resulted from the fact that a non-local $S^1$ along $\hat{x}_1$ must be the intersection of two perpendicular $M^2$'s respectively of $\langle \hat{x}_1, \hat{x}_2 \rangle$ and $\langle \hat{x}_1, \hat{x}_3 \rangle$ directions. So the consistent condition between $x_{M^2}$'s is that we can only change the $\mathsf{SGS}$'s on such two perpendicular $M^2$'s simultaneously. Due to that, for a X-cube model of the size $L\times L \times L$ with periodic boundary condition (PBC), the characteristic graph of the above mentioned part is just a square lattice of the size $L\times L$ with PBC. An example of the characteristic graph of X-cube model and its correspondence with ground states is given in Fig.~\ref{fig:charac}.

\begin{figure}[t]
	\includegraphics[width=0.46\textwidth]{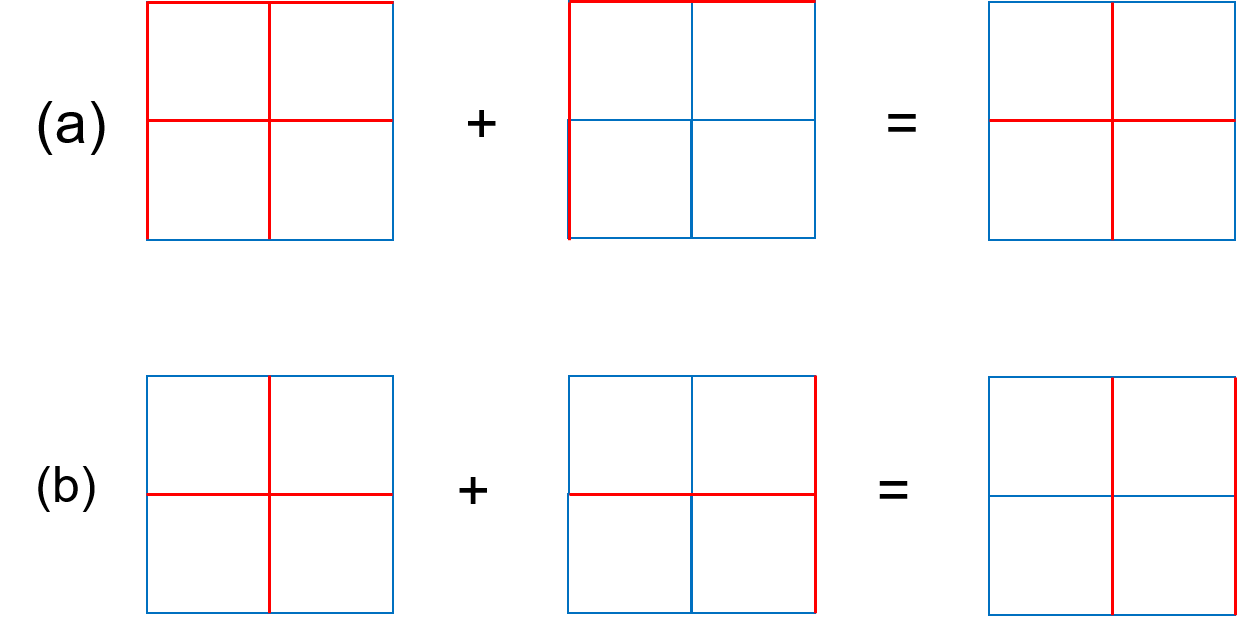}
	\caption{\textbf{Examples of the multiplication of edge-coloring patterns.} Here we only draw the graph of the size $3\times 3$.}
	\label{fig:multiplication}
\end{figure}

Furthermore, in order to count the number of possible edge coloring patterns, we introduce a ``characteristic group'' based on following rules:
\begin{itemize}
	
	\item An edge-coloring pattern of a characteristic graph is recognized as a group element.
	
	\item A vertex in the characteristic graph corresponds to a generator of the group, which is the edge-coloring pattern obtained by coloring all lines that intersect at the vertex. Different vertices correspond to different generators.
	
	\item The multiplication result of two elements is the pattern generated by the product of the corresponding $W(S^{d_s})$ operators. We can see the multiplication is equivalent to $\Z_2$ addition of coloring of lines. Fig.~\ref{fig:multiplication} gives an example of such a multiplication in X-cube model.
	
	\item Every generator defined above is its own inverse element.
	
	\item The completely uncolored pattern is the identity element.
	
\end{itemize}

As the identity elements and inverse elements are already given in the above rules, we only need to check whether the above defined structure is closed and associative. According to our definition of the multiplication, all multiplication results can be generated by vertex generators, so the structure is closed. And since the multiplication is equivalent to the $\Z_2$ addition of coloring of lines, the multiplication is associative and commutative. Therefore, we can see that the structure is an Abelian group. 

For an $n$-dimensional characteristic graph, the product of the vertex generators which form an $n$-cube equals to the identity. That is to say, not all vertex generators are independent. We need to find the number of independent generators to obtain the order of the characteristic group. 

\begin{figure}[t]
	\includegraphics[width=0.45\textwidth]{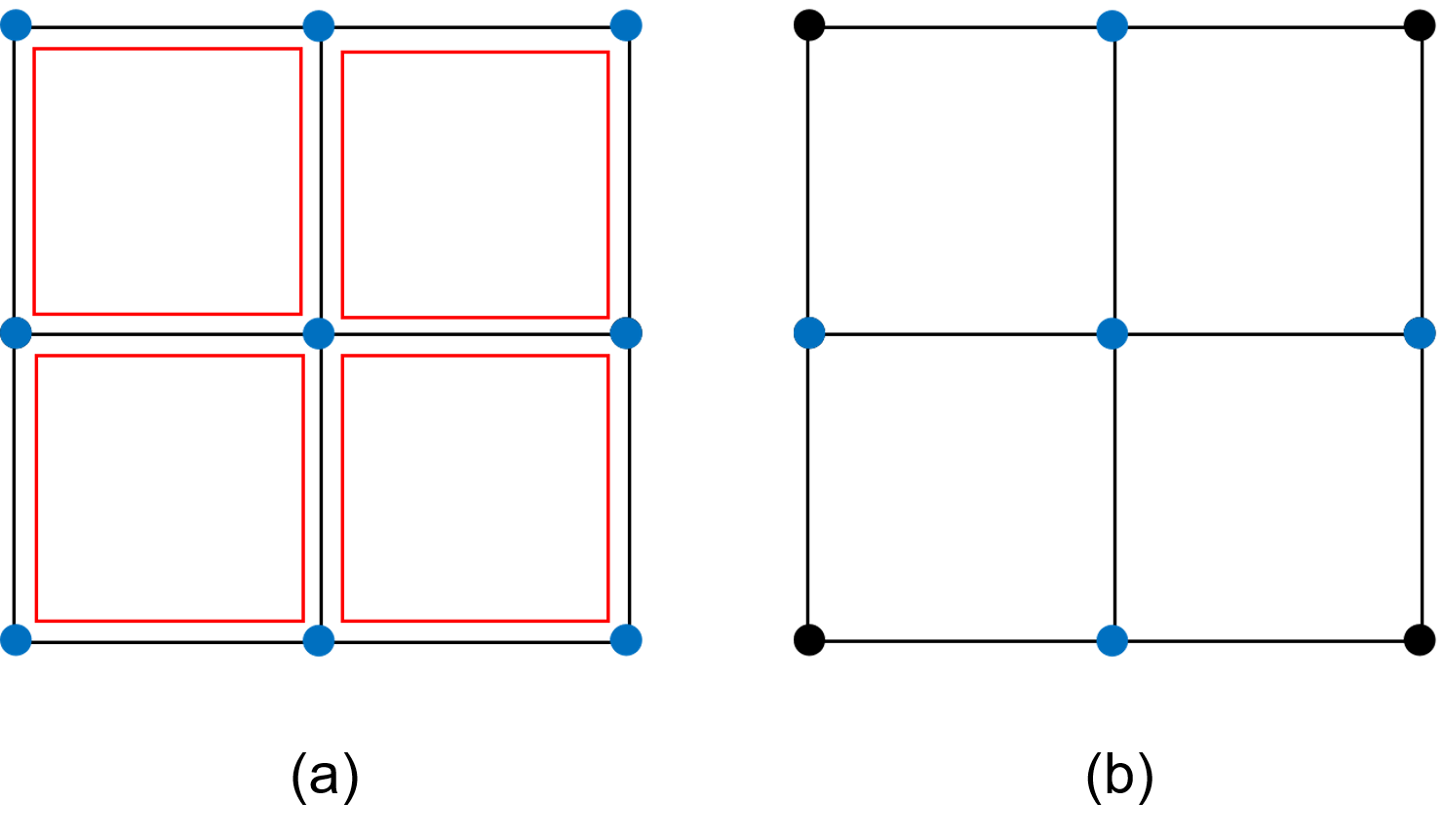}
	\caption{\textbf{Pictorial demonstration of the elimination of additional degrees of freedom in the characteristic graphs.} The red squares in (a) represent the independent constraints associated with plaquettes, and the blue vertices represent the generators. In (b), since each independent constraint makes one generator in a plaquette redundant, we can simultaneously forget the redundant generators and the constraints to get $5$ independent degrees of freedom.}
	\label{fig:constraint}
\end{figure}

In the X-cube case, we continue with the discussion of characteristic graph of the $\hat{x}_1$ part, which is a square lattice of the size $L\times L$ with PBC. If we ignore the constraints given by $2$-cubes in the graph, the characteristic group has $L^2$ vertex generators, so it should be a direct product of $L^2$ $\Z_2$ groups. When constraints are taken into consideration, we notice that only $(L-1)^2$ constraints that correspond to the $\gamma_2$'s (i.e. plaquettes)\footnote{Apparently, there are $L^2$ $\gamma_2$'s in the characteristic graph, but we can verify that the extra $\gamma_2$'s resulted from the PBC can always be generated only by the $\gamma_2$'s that exist with OBC. It is a general rule due to the correspondence between vertices and generators.} in the characteristic graph are necessary to generate all constraints. We can use each independent constraint to eliminate one generator in a plaquette. At last, only $L^2-(L-1)^2=2L-1$ independent $\Z_2$ degrees of freedom remain, so the group has $2^{2L-1}$ elements in total (Fig.~\ref{fig:constraint} gives a pictorial demonstration of the elimination of degrees of freedom). Therefore, the ground state degeneracy of a X-cube model defined on a lattice of the size $L\times L\times L$ with PBC should be $(2^{2L-1})^3=2^{6L-3}$ (since we have dismantled the ground states into $3$ independent parts), that is to say, we have $\log_2 GSD=6L-3$, which is consistent with the known result \cite{Shirley2018}. 

 {In conclusion, once we have determined the $\mathsf{SGS}$'s and consistent conditions of a $[D-3,D-2,D-1,D]$ or $[0,1,2,D]$ model, we can then write down a series of characteristic graphs corresponding to ground states of different parts, and obtain the GSD by computing the orders of characteristic groups. }

\section{Typical examples}
\label{sec_examples}

\subsection{GSD of $[0,1,2,D]$ models on isotropic lattices}
\label{subsec:012D_models}

We begin with the discussion about $[0,1,2,D]$ models of the size $\underbrace{L\times L\times ... \times L}_{D\text{ times}}$ with PBC. 
\begin{itemize}
\item First, we analyze the ground states of a $[0,1,2,D]$ model to identity the $\mathsf{SGS}$'s. As a stabilizer code model, a ground state $|\phi \ket$ of a $[0,1,2,D]$ model has to satisfy the following conditions:
\begin{align*}
B^l_{\gamma_0} |\phi \ket &= |\phi \ket,\ \forall \gamma_0,l;\\
A_{\gamma_D} |\phi \ket &= |\phi \ket,\ \forall \gamma_D.
\end{align*}
In $\sigma^z$ basis, conditions given by $B$ terms require in ground states, all Ising configurations $c$ can only have $S^1$ (i.e. straight strings) and vertices emanating $D$ perpendicular $S^1$'s. And conditions given by $A$ terms require all Ising configuration $c$'s that can be connected by action of $A$ terms to belong to the same ground state.

For $\mathsf{SGS}$'s on $M^2$, we can see that subsystem Ising configurations $c_{M^2}$'s can only have $S^1$ and vertices emanating $2$ perpendicular $S^1$'s, that is to say, $[\rho^{\mathrm{c}}]^{M^D}_{M^2}(c)=c_{M^2}$ can only have closed loops. And an $\mathsf{SGS}$ is composed of all such $c_{M^2}$'s that can be connected by $[\rho^{\mathrm{o}}]^{M^D}_{M^2}(A_{\gamma_D})=A_{\gamma_2}=\prod_{\gamma_1\in \gamma_2} \sigma^x_{\gamma_1}$ operators. In summary, an $\mathsf{SGS}$ on an $M^2$ is equivalent to a ground state of a 2D toric code model. Different $\mathsf{SGS}$'s can be distinguished by the action of non-local $W(S^1) = \prod_{\gamma_1 \in S^1} \sigma^x_{\gamma_1}$ operators.

\item Second, we consider the consistent conditions between these $\mathsf{SGS}$'s. Since $W(S^1)$'s with $S^1$ along different directions contribute to $\mathsf{SGS}$'s independently, without loss of generality, now we restrict our discussion to $S^1$'s along $\hat{x}_1$ direction. As discussed in Sec.~\ref{subsec:overview}, the consistent conditions requires a $W(S^1)$ operator to simultaneously act on $(D-1)$ intersecting $M^2$'s. For $S^1$'s along $\hat{x}_1$ direction, the $(D-1)$ $M^2$'s must be respectively of the $\langle \hat{x}_1, \hat{x}_2\rangle$, $\langle \hat{x}_1, \hat{x}_3\rangle$, $\langle \hat{x}_1, \hat{x}_4\rangle$, $\cdots$, and $\langle \hat{x}_1, \hat{x}_D\rangle$ directions. 

\item Third, we summarize the above data of $\mathsf{SGS}$'s and their consistent conditions in the characteristic graph, and obtain the GSD from the graph. For the part with $S^1$'s along $\hat{x}_1$ direction, the characteristic graph is a $(D-1)$-dimensional hypercubic lattice with PBC. So according to our discussion in Sec.~\ref{subsec:overview}, the characteristic group has $L^{(D-1)}$ generators (corresponding to the vertices in the graph) and $(L-1)^{(D-1)}$ constraints (corresponding to the $(D-1)$-cubes in the graph). Because the system we are considering is isomorphic, all the $D$ parts give the same results, and the final GSD is just given by the product of the results of different parts. 
 
\end{itemize}

Finally, we obtain that the GSD of $[0,1,2,D]$ models is given by:
\begin{align}
\log_2 GSD=&D\times (L^{D-1}-(L-1)^{D-1})\nonumber\\
=& D \times \sum_{n=0}^{D-2}  \binom{D-1}{n} (-1)^{D+n} L^n\,.
\end{align}
When $D=3$, the result automatically restores to the X-cube case.
\subsection{GSD of $[0,1,2,D]$ models on anisotropic lattices}

For a $[0,1,2,D]$ model defined on a lattice of the size $\underbrace{L_1\times L_2 \times L_3 \times \cdots \times L_D}_{D\text{ times}}$ with PBC, a ground state follows exactly the same manner as in the isotropic case. That is to say, the $\mathsf{SGS}$'s are also equivalent to 2D toric code ground states. Besides, the consistent conditions are also the same. The anisotropic condition only influence the sizes of characteristic graphs.

For the part of $S^1$'s along $\hat{x}_i$ ($1\leq i\leq D$) direction, the characteristic graph is a $(D-1)$-dimensional hypercubic lattice of the size $L_1\times L_2 \times L_3 \times \cdots \times \hat{L_i} \times \cdots \times L_D$. Here $\hat{L_i}$ means that $L_i$ is not included in the product. Therefore, the corresponding characteristic group has $L_1\times L_2 \times L_3 \times \cdots \times \hat{L_i} \times \cdots \times L_D$ generators and $(L_1-1)\times (L_2-1) \times (L_3-1) \times \cdots \times \hat{(L_i-1)} \times \cdots \times (L_D-1)$ independent constraints. 

As the contribution of other parts can be similarly obtained, the GSD of a $[0,1,2,D]$ model defined on a anisotropic lattice is given by:
\begin{align*}
\nonumber
\log_2 GSD
=&\sum_j^D \{(\frac{1}{L_j} \prod_i^D L_i) - (\frac{1}{L_j-1} \prod_i^D (L_i-1)) \} \\
=&\ 2 \times (\sum_{k_1}^D \sum_{k_2}^{k_1 - 1} \frac{1}{L_{k_1} L_{k_2}} \prod_n L_n) \\
&-3 \times (\sum_{k_1}^D \sum_{k_2}^{k_1 - 1} \sum_{k_3}^{k_2 - 1} \frac{1}{L_{k_1} L_{k_2} L_{k_3}} \prod_n L_n) \\
&+4 \times (\sum_{k_1}^D \sum_{k_2}^{k_1 - 1} \sum_{k_3}^{k_2 - 1} \sum_{k_4}^{k_3 - 1}\frac{1}{L_{k_1} L_{k_2} L_{k_3} L_{k_4}} \prod_n L_n)\\
&- \ \cdots \\
&+(-1)^{D-1} (D-1) \sum_{n}^{D} L_n\\
&+(-1)^D D. 
\end{align*}

\subsection{GSD of $[D-3,D-2,D-1,D]$ models on isotropic lattices}
\label{subsec:-3-2-1D_models}

Then we consider $[D-3,D-2,D-1,D]$ models of the size $\underbrace{L\times L\times ... \times L}_{D\text{ times}}$ with PBC. 

\begin{itemize}
	
	\item First, we analyze the ground states of a $[D-3,D-2,D-1,D]$ model to identity the $\mathsf{SGS}$'s. As a $[D-3,D-2,D-1,D]$ model is also a stabilizer code, a ground state $|\phi \ket$ of a $[D-3,D-2,D-1,D]$ model analogously satisfy the following conditions:
	\begin{align*}
	B^l_{\gamma_{D-3}} |\phi \ket &= |\phi \ket,\ \forall \gamma_{D-3},l;\\
	A_{\gamma_D} |\phi \ket &= |\phi \ket,\ \forall \gamma_D.
	\end{align*}
	Here we should notice that the definition of $A_{\gamma_D}$ terms are different from the $A$ terms in a $[0,1,2,D]$ models, though they share the same form (see Sec.~\ref{subsec:review}). In $\sigma^z$ basis, conditions given by $B$ terms require that in ground states, all Ising configuration $c$'s can only have $S^{D-2}$'s and $S^{D-3}$'s emanating $3$ perpendicular $S^{D-2}$'s. And conditions given by $A$ terms require all $c$'s that can be connected by action of $A$ terms to belong to the same ground state.
	
	For $\mathsf{SGS}$'s on $M^{D-1}$, we can see that subsystem Ising configurations ${c}_{M^{D-1}}$'s can only have $S^{D-2}$'s and $S^{D-3}$'s emanating 2 perpendicular $S^{D-2}$'s, that is to say, $[\rho^{\mathrm{c}}]^{M^D}_{M^{D-1}}(c)={c}_{M^{D-1}}$ can only have closed $(D-2)$-dimensional objects. And an $\mathsf{SGS}$ is composed of all such ${c}_{M^{D-1}}$'s that can be connected by $[\rho^{\mathrm{o}}]^{M^D}_{M^{D-1}}(A_{\gamma_D})=A_{\gamma_{D-1}}$ operators. Different $\mathsf{SGS}$'s can be distinguished by the action of non-local $W(S^{D-2}) = \prod_{\gamma_{D-2} \in S^{D-2}} \sigma^x_{\gamma_{D-2}}$ operators, where $S^{D-2}$ is closed.
	
	\item Second, we consider the consistent conditions between these $\mathsf{SGS}$'s on $M^{D-1}$'s. Again, $W(S^{D-2})$'s with $S^{D-2}$ along different directions contribute to $\mathsf{SGS}$'s independently. Without loss of generality, we restrict our discussion to $S^{D-2}$'s along $\langle \hat{x}_1, \hat{x}_2,\hat{x}_3,\cdots,\hat{x}_{D-2}\rangle$ direction. As discussed in Sec.~\ref{subsec:overview}, the consistent condition requires a $W(S^{D-2})$ operator to simultaneously act on $2$ intersecting $M^{D-1}$'s. For $S^{D-2}$'s along $\langle \hat{x}_1, \hat{x}_2,\hat{x}_3,\cdots,\hat{x}_{D-2}\rangle$ direction, the $2$ $M^{D-1}$'s must be respectively of the $\langle \hat{x}_1, \hat{x}_2,\hat{x}_3,\cdots,\hat{x}_{D-2},\hat{x}_{D-1}\rangle$ and $\langle \hat{x}_1, \hat{x}_2,\hat{x}_3,\cdots,\hat{x}_{D-2},\hat{x}_{D}\rangle$ directions. 
	
	\item Third, we summarize the above data of $\mathsf{SGS}$'s and their consistent conditions in the characteristic graph, and obtain the GSD from the graph. For the part with $S^{D-2}$'s along $\langle \hat{x}_1, \hat{x}_2,\hat{x}_3,\cdots,\hat{x}_{D-2}\rangle$ direction, the characteristic graph is a $2$-dimensional square lattice with PBC. So according to our discussion in Sec.~\ref{subsec:overview}, the characteristic group has $L^2$ generators (corresponding to the vertices in the graph) and $(L-1)^2$ constraints (corresponding to the $2$-cubes in the graph). Because the system we are considering is isomorphic, all the $\binom{D}{D-2}$ parts give the same results, and the final GSD is just given by the product of the results of different parts. 
	
\end{itemize}

Finally, we obtain that the GSD of $[D-3,D-2,D-1,D]$ models is given by:
\begin{align}
\log_2 GSD=&\binom{D}{D-2} \times (L^2-(L-1)^2)\nonumber\\
=&\binom{D}{D-2} \times (2L-1)\\
=&(D^2-D) \times L -\frac{D^2-D}{2}.
\end{align}
Again, when $D=3$, the result will restore to the X-cube case.

\subsection{GSD of $[D-3,D-2,D-1,D]$ models on anisotropic lattices}
\label{subsec:ani_lat}

The difference between a $[D-3,D-2,D-1,D]$ model defined on a lattice of the size $\underbrace{L_1\times L_2 \times L_3 \times \cdots \times L_D}_{D\text{ times}}$ and the isotropic case is similar to a $[0,1,2,D]$ model. All $\mathsf{SGS}$'s and their consistent conditions stay the same, we only need to consider variance of the characteristic graphs.

For the part of $S^{D-2}$'s along $\langle \hat{x}_1, \hat{x}_2,\hat{x}_3,\cdots,\hat{x}_{D-2}\rangle$ direction, the characteristic graph is a $2$-dimensional square lattice with the size $L_{D-1}\times L_D$. Therefore, the corresponding characteristic group has $L_{D-1} \times L_D$ generators and $(L_{D-1}-1) \times (L_D-1)$ independent constraints. 

As the contribution of other parts can be similarly obtained, the GSD of a $[D-3,D-2,D-1,D]$ model defined on an anisotropic lattice is given by:
\begin{align*}
\log_2 GSD=& \sum_i^D \sum_j^{i-1} \{L_i \times L_j - (L_i - 1)\times (L_j - 1)\}\\
=&\sum_i^D \sum_j^{i-1} \{L_i + L_j - 1\}\\
=&\sum_i^D (D-1)\times L_i-\frac{D^2-D}{2}.
\end{align*}

\section{Discussions and Conclusions}
\label{sec_conc}

By comparing the GSD of X-cube, $[0,1,2,D]$ and $[D-3,D-2,D-1,D]$ models summarized in Table~\ref{table_result}, we have the following observations. First, in a $[0,1,2,D]$ model, the $\log_2 GSD$ becomes a polynomial. Such a polynomial contains not only the linear and constant terms, but also terms of any degrees that  are smaller than $D-1$. Besides, in a $[D-3,D-2,D-1,D]$ model, the $\log_2 GSD$ is of same form as X-cube model. 

The polynomials $\log_2 GSD$ in $[0,1,2,D]$ models are the most interesting and unprecedented result here, while a complete understanding of these polynomials is still lacking. Here we believe it is beneficial to give some primary discussions about the relation between polynomials $\log_2 GSD$ and other features of fracton orders.

First, as we have mentioned in Sec.~\ref{sec:intro}, in $[0,1,2,D]$ models, from the perspective of foliated fracton orders, their leaves  admit foliation structures as well. Therefore, we may recognize the orders in $[0,1,2,D]$ models as multi-level foliated fracton orders. Or from the perspective of decomposition, it means that when the whole system is in a ground state, the $\mathsf{SGS}$ in a $(D-1)$-dimensional subsystem also looks like a ground state of a fracton model (more exactly, it looks like a ground state of a $[0,1,2,D-1]$ model). Mathematically,  we have already known the relation between the coefficients in the $\log_2 GSD$ of X-cube model and the $2$-foliation of the manifold where the system is defined \cite{Shirley2018}. The coefficients in all polynomials found in this work are expected to reflect a series of topological and geometric properties of the quantum models and background lattice. For example, as the multi-level foliation structure suggests, the coefficients are likely to contain information of subsystems of various dimensions. Now, as a rigorous and complete discussion of multi-level foliation is still under exploration, some future works seem to be necessary to clarify the relationship between models like $[1,2,3,D]$ and multi-level foliation structures.

Second, in our previous work,   topological excitations are classified into \textit{trivial excitations} (denoted as $\mathbb{I}$), simple excitations (denoted as $\mathsf{E}^s$), \textit{complex excitations}  (denoted as $\mathsf{E}^c$) and \textit{intrinsically disconnected excitations} (denoted as $\mathsf{E}^d$)   in $[d_n,d_s,d_l,D]$ models \cite{ye19a}, according to mobility and deformability of topological excitations. Fracton physics of spatially extended excitations can exist in all nontrivial excitations, i.e., $\mathsf{E}^s$, $\mathsf{E}^c$ and $\mathsf{E}^d$, as studied in Ref.~\cite{ye19a}. While the GSD polynomials are also derived from   $[d_n,d_s,d_l,D]$ models,  one may wonder what is the relation between GSD polynomials and existence of nontrivial excitations.  A similar question can be answered  in (2+1) Abelian topological order where GSD on a torus  is equal to the number of distinct topological excitations. But it is hard to firmly answer  the question in the present fracton models, which is left to future work.
  

Besides, we believe the mathematical aspect of polynomial $\log_2 GSD$ may be particularly interesting. Inspired from the foliated fracton orders, we believe the coefficients of such polynomials may reflect the properties of various subsystems. While if we consider such a polynomial as a whole, the polynomial itself may represent a novel characteristic of   topology and geometry of a high-dimensional object. That is to say, how these coefficients interact each other, and how these information of different subsystems are integrated into a whole, may require more mathematical exploitation. These questions may be answered by considering $[d_n,d_s,d_l,D]$ models on general manifolds \cite{Shirley2018} which are beyond toric manifolds used in this work.

From another perspective, we can see that the relation between $[0,1,2,4]$ and X-cube model is similar to the relation between X-cube and 2D toric code model. It makes $[0,1,2,4]$ model kind of a ``fracton order of fracton order''. Such an observation makes us believe that better knowledge about the relation between higher-dimensional fracton orders and 3D fracton orders may be beneficial to gain deeper insights about a general theory of fracton orders and especially the physics of non-manifold excitations found in Ref.~\cite{ye19a}.

In this paper, as we have restricted our discussion to models with $[d_n,d_s,d_l,D]$ models with $d_s-d_n=1$, the generality of computing GSD based on SGS's is still kind of vague. If it is possible to generalize our computation to other foliated fracton orders, or even Type-II fracton orders, like Haah's code, is quite worth exploring. Nonetheless, until now we can only case-by-case check if the SGS method works for a given model. Besides, as in principle, we need to exhaust all manifold decomposition $\mathcal{M}$ to assert the SGS method does not work for a model, it seems difficult to understand when the SGS method would fail. 

In order to compute the GSD, we demonstrate a method to represent a ground state with a collection of lower-dimensional data. Then it is a natural question to ask if such lower-dimensional representations can be applied to excited states. Since excitations can have representations as composites of collections of subsystem superselection sectors with the form of $(e\ m)+(m\ e)$, such composites may be recognized as being composed of ``entangled'' topological excitations. Therefore, we believe the decomposition of excitations may also reveal some interesting features of fracton orders, which will be presented in  \cite{futureLiYe2021}.

\section*{Acknowledgements}
This work was supported in part by the Sun Yat-sen University startup grant, Guangdong Basic and Applied Basic Research Foundation under Grant No. 2020B1515120100, National Natural Science Foundation of China (NSFC) Grant (No. 11847608 \& No. 12074438).

\appendix

\section{Notations and conventions}

\begin{itemize}
	
	\item{$\gamma_d$}: d-cube. Firstly appears in Sec.~\ref{subsec_geo_not}.
	
	\item{$S^d$}: a $d$-dimensional analog of a cuboid. Firstly appears in Sec.~\ref{subsec_geo_not}.
	
	\item{$M^d$}: a $d$-dimensional manifold. Firstly appears in Sec.~\ref{subsec_res_of_gs}.
	
	\item{$\mathcal{M}$}: a $(d_1,d_2,\cdots,d_k)$-decomposition of a given $D$-dimensional base manifold $M^D$. Firstly appears in Sec.~\ref{subsec_res_of_gs}.

	\item{$[\rho^{\mathrm{c}}]^{M^{d_2}_j}_{M^{d_1}_i}(c_{M^{d_2}_j})$}: a restriction of an Ising configuration on $M^{d_2}_j$ to one of its subsystem $M^{d_1}_i$. Firstly appears in Sec.~\ref{subsec_res_of_gs}.
		
	\item{$\Omega^{\mathrm{c}}_{\mathcal{M}}(c)$}: a decomposition of an Ising configuration $c$ on the base manifold to a collection of subsystem specified by $\mathcal{M}$. Firstly appears in Sec.~\ref{subsec_res_of_gs}.
	
	\item{$[\rho^{\mathrm{o}}]^{M^{d_2}_j}_{M^{d_1}_i}(o_{M^{d_2}_j})$}: a restriction of an X-operator on $M^{d_2}_j$ to one of its subsystem $M^{d_1}_i$. Firstly appears in Sec.~\ref{subsec_res_of_gs}.
	
	\item{$\Omega^{\mathrm{o}}_{\mathcal{M}}(o)$}: a decomposition of an X-operator $o$ on the base manifold to a collection of subsystem specified by $\mathcal{M}$. Firstly appears in Sec.~\ref{subsec_res_of_gs}.
	
	\item{$c_M$}: an Ising configuration $c$ on manifold $M$. When the manifold can be specified from the context, we may omit the subscript $M$.
	
	\item{$[\rho^{\mathrm{x}}]^{M^D}_{M^{d_1}_i}(x)$}: a restriction of a ground state $x$ on the base manifold $M^D$ to one of its subsystem $M^{d_1}_i$. Firstly appears in Sec.~\ref{subsec_res_of_gs}.
	
	\item{$\Omega^{\mathrm{o}}_{\mathcal{M}}(o)$}: a decomposition of an a ground state $x$ on the base manifold to a collection of subsystem specified by $\mathcal{M}$. Firstly appears in Sec.~\ref{subsec_res_of_gs}.
	
	\item{GSD}: ground state degeneracy. Firstly appears in Sec.~\ref{sec:intro}.
	
	\item{$\mathsf{SGS}$}: subsystem ground state sector. Firstly appears in Sec.~\ref{subsec_res_of_gs}.
	
	\item{PBC}: periodic boundary condition. Firstly appears in Sec.~\ref{subsec:overview}.
	
	\item{OBC}: open boundary condition. Firstly appears in Sec.~\ref{subsec:overview}.
		
\end{itemize}
	
\section{Proof of the isomorphism between ground states and consistent combinations of $\mathsf{SGS}$'s}
\label{sec_proof}

In this Appendix, we prove that for a $[d_n,d_s,d_l,D]$ model defined on toric $M$ with a manifold decomposition $\mathcal{M}$, given that (a) $\mathcal{M}$ contains $(d_s+1)$-dimensional subsystems and (b) $d_s-d_n=1$, then there is always an isomorphism between ground states of a $[d_n,d_s,d_l,D]$ model and consistent combinations of $\mathsf{SGS}$'s on $\mathcal{M}$. That is to say, if a combination of $\mathsf{SGS}$'s can be obtained as a decomposition of a ground state $x$ on $\mathcal{M}$, then we can always obtain the original ground state $x$ merely from the information of these $\mathsf{SGS}$'s.
\begin{proof}
	We use contradiction to prove it. For a given $\mathcal{M}$ including $\{M^{d_s+1}\}$, we assume that there exists a pair of ground states $x_1\neq x_2$, such that 
	\begin{align}
	\label{eq_a2_1}
	\Omega^{\mathrm{x}}_{\mathcal{M}}(x_1) = \Omega^{\mathrm{x}}_{\mathcal{M}}(x_2).
	\end{align} 
	As $x_1$ and $x_2$ are different ground states, they can always be connected by a logical operator denoted as $o_x$. So we have \begin{align}
	\label{eq_a2_2}
	o_x|x_1\rangle=|x_2\rangle. 
	\end{align} 
	
	In a $[d_n,d_s,d_l,D]$ model, when $d_s-d_n=1$, we can consider a complete set of logical operators generated by $W(S^{d_s})=\prod_{\gamma_{d_s}\in S^{d_s}} \sigma^x_{\gamma_{d_s}}$, where $S^{d_s}$ is closed. These logical operators are composed of $\sigma^x$ operators along closed $S^{d_s}$'s. Here, as an $S^{d_s}$ is required to be totally flat, a closed $S^{d_s}$ must be extended along all $d_s$ directions to avoid any boundaries. 
	
	Then, we need to notice not all such $W(S^{d_s})$ generators are independent to each other. The product of $2^{D-d_s}$ $W(S^{d_s})$'s equals to the trivial logical operator, when these $S^{d_s}$' form the $d_s$-dimensional boundary of an $S^D$ that is extended along $d_s$ directions. That is to say, these $S^{d_s}$' satisfy the following conditions:
	\begin{itemize} 
		\item All $S^{d_s}$'s are parallel;
		\item There are only $2$ or $0$ $S^{d_s}$'s in an $M^{d_s+1}$.
	\end{itemize}
	Such a product is trivialized as it can be written as a product of $A_{\gamma_D}$ operators. 
	
	As $o_x$ preserves all $\mathsf{SGS}$'s, from Eq.~\ref{eq_a2_1}, we can see that $\forall\ M^{d_s+1} \in \mathcal{M}$, $[\rho^{\mathrm{o}}]^M_{M^{d_s+1}}(o_x)$ must be a product of parallel pairs of $W(S^{d_s})$ operators. Therefore, $\forall\ M^{d_s+2} \subset M$, $[\rho^{\mathrm{o}}]^M_{M^{d_s+2}}(o_x)$ must be a product of parallel quadruples of $W(S^{d_s})$ operators. Iteratively, we obtain that $o_x$ must be the product of groups of $2^{D-d_s}$ parallel $W(S^{d_s})$ operators. Consequently, on a toric base manifold $M$, by removing $W(S^{d_s})$ operators which compose trivial logical operators, $o_x$ can be transformed to the identity operator, whose restrictions on all $M^{d_s+1}$'s are trivial. As a result, $o_x$ has to be the trivial logical operator, and it is against our assumption that $x_1\neq x_2$. 
\end{proof}
Thereupon, the map from ground states to combinations of subsystem ground state sectors is an injection. So a ground state $x$ of a $[d_n,d_s,d_l,D]$ model can be reconstructed from its decomposition on $\mathcal{M}$. 

Finally, we make some comments about the case when $M$ is not a torus, but a manifold with more complicated topology. Then, the above proof may fail as even though $[\rho^{\mathrm{o}}]^M_{M^{d_s+1}}(o_x)$ is a product of $2$ $W(S^{d_s})$ operators for an arbitrary $M^{d_s+1} \in \mathcal{M}$, it may be impossible to find a product of $A_{\gamma_D}$ operators to reduce all these $W(S^{d_s})$ operators simultaneously. In that case, to reconstruct the original ground state, except for $\mathsf{SGS}$'s, we may also need additional information depending on the topology of the base manifold $M$. Nevertheless, it is also possible that the original ground state can be reconstructed by adding new subsystems to manifold decomposition $\mathcal{M}$. As far as we know, more works are needed to obtain a systematic understanding of such problems.


%


\end{document}